\documentclass[journal]{IEEEtran}
\usepackage{amsmath,amsthm,amssymb}
\usepackage{algorithm,algorithmic}
\usepackage{xspace}
\usepackage{url}
\usepackage{caption}
\usepackage{subcaption}
\usepackage{enumerate}
\usepackage{graphicx,epsfig}

\begin{document}
%
\title{Topological and Statistical Behavior Classifiers for Tracking Applications}
%
%
%

\author{
    \IEEEauthorblockN{Paul Bendich\IEEEauthorrefmark{1}, Sang Chin\IEEEauthorrefmark{2}, Jesse Clarke\IEEEauthorrefmark{2}, Jonathan deSena\IEEEauthorrefmark{2}, John Harer \IEEEauthorrefmark{1}, Elizabeth Munch \IEEEauthorrefmark{1}, Andrew Newman \IEEEauthorrefmark{2}, David Porter \IEEEauthorrefmark{2}, David Rouse \IEEEauthorrefmark{2}, Nate Strawn \IEEEauthorrefmark{2}, Adam Watkins \IEEEauthorrefmark{2}}
    
    \IEEEauthorblockA{\IEEEauthorrefmark{1} Departments of Mathematics, Computer Science, and Electrical and Computer Engineering, Duke University, Durham, NC 27708}
    
    \IEEEauthorblockA{\IEEEauthorrefmark{2}Johns Hopkins University Applied Physics Laboratory, Laurel, MD, 20723}
}

\maketitle

\begin{abstract}
We introduce the first unified theory for target tracking using Multiple Hypothesis Tracking, Topological Data Analysis, and machine learning. Our string of innovations are 1) robust topological features are used to encode behavioral information, 2) statistical models are fitted to distributions over these topological features, and 3) the target type classification methods of Wigren and Bar Shalom et al. are employed to exploit the resulting likelihoods for topological features inside of the tracking procedure. To demonstrate the efficacy of our approach, we test our procedure on synthetic vehicular data generated by the Simulation of Urban Mobility package. 
\end{abstract}


%
\IEEEpeerreviewmaketitle

\section{Introduction}

\label{sec:introduction}

\IEEEPARstart{T}{here} has been a growing awareness in the tracking and data fusion communities that target 
behavior provides key information for enabling successful tracking under challenging conditions, 
particularly when success is based upon Activity-Based Intelligence (ABI). 
As sensor capabilities and tracking algorithms are employed to achieve real-time, wide area, 
multi-sensor tracking with confidence, the information that can be provided by target activity 
and pattern-of-life provide exciting new dimensions to meet these challenges. 
Furthermore, as tracking methods are extended to the higher-level fusion problems (such as situation 
awareness and determination of intent) behavior estimation becomes particularly important. 

In this paper, we present the first unified theory for target tracking using Multiple Hypothesis Tracking (MHT), 
Topological Data Analysis (TDA), and machine learning. 
The key idea is to use topologically-inspired measures of behavior to prune out improbable ``tracklets" inside of the MHT. 
TDA not only provides classification of target behavior, but (perhaps even more importantly) also 
helps solve what is often the most difficult problem in target tracking: connecting-the-dots 
by associating data with targets. 
Intuitively, data are associated in a manner not only consistent relative to target location-related data 
and type-related data, but also consistently relative to behavior-related data. 
For example, tracking urban traffic with Wide Area Motion Imagery (WAMI) data may 
benefit more from distinguishing vehicles by their behavior than from trying to distinguish 
vehicles using fuzzy imagery.

This paper uses topology to characterize agent behavior by simple methodology, and we demonstrate the value of our approach using simulated urban traffic observed by a WAMI sensor system. In addition, we believe that the results presented here only scratch the surface of our potential contributions as our basic ideas immediately penetrate into areas such as coordinated behavior of groups of targets, behavior of components of extended target systems that may not be collocated (such as an air defense system), 
and even behavior exploitable by cyber data fusion.

\subsection{Multiple Hypothesis Tracking.}

Most modern multi-target multi-sensor tracking systems use some version of the MHT framework pioneered by 
Reid \cite{Reid1979}, and comprehensively described in Blackman and Popoli \cite{Blackman1999}
and by Hall and Llinas \cite{Llinas2009}. 
For the last decade or two, the dominant approaches solve a Bayesian inverse probability 
problem that scores competing multi-track hypotheses with a Bayesian Log-Likelihood 
Ratio (LLR) as described by Bar Shalom and collaborators \cite{BarShalom2007}.
Data are processed recursively with a deferred decision logic scheme that expands 
parent hypotheses with new data to form child hypotheses. These child hypothesis are subsequently pruned to provide 
a reasonable number of parent hypotheses for the next round of data. 
The track file of pruned candidate tracklets can become much larger than the actual number of
targets, where the tracklets are alternative attempts to represent targets with the same data. 

For urban traffic with WAMI data, images from multiple cameras are processed using a motion 
detection algorithm to segment pixels into candidate vehicles. 
As vehicles slow, particularly at intersections, motion detection becomes difficult. Further, algorithms for 
detecting stationary vehicles are much less effective. 
Consequently, the tracklets can be quite confused at intersections. 
Often there will be tracklets trying to represent the same vehicle that make all possible turning decisions 
at the intersection. In other words, it can be difficult to connect-the-dots with fuzzy images of vehicles that 
are near each other. 
However, if the behavior for a tracklet after the intersection is similar to behavior before the intersection, 
then it would seem the score of the tracklet should be increased and that the confusion could be resolved. 
This is in fact what the behavior flags and TDA coupled with machine learning accomplish as described 
below.

The unified approach taken in this paper is motivated by a previous extension of MHT to include target type classification 
by Wigren \cite{Wigren2001} and Bar Shalom, et alia \cite{BarShalom2005}.
Many attempts had been made to incorporate target type data in MHT, but most of them were ad hoc. 
Wigren and Bar Shalom provided a unified approach that used target type data to compute an additive 
term for the track LLR. 
Not only was target type classification provided, but data association was improved. 
In fact, this was an enabling technology for field-scale implementation of Upstream Data Fusion 
(UDF) methods reported by Newman and Mitzel \cite{Newman2013}.

Target behavior classification is incorporated into MHT in a similar manner as target type classification. 
A difference is that target behavior uses a window of data over which the behavior can be observed,
whereas target type can be observed based only on current data. 
Consequently, the target behavior algorithm is multi-rate with the target kinematic state estimation and 
target type classification running at a faster rate than the target behavior classification. 
Another point is that both target type and behavior can be incorporated treating classification with 
current observations as the data, or treating the upstream features used by the classifier as the data.

\subsection{Topological Data Analysis.}
To each tracklet we want to associate functions that describe behavior.
We focus on speed, acceleration and turning, 
but the general framework will work for any functions.
Figure \ref{fig:speeds}  illustrates speed functions of different behavior types.
Given two tracklets $T_1$ and $T_2$, we will use these functions to help us figure out whether or 
not the agents associated to $T_1$ and $T_2$ are the same.

There are of course many different ways to summarize functional data: critical values, total variation, motif-hunting,
to name a few.
Our proposed solution is to use a \emph{persistence diagram} (PD),
which we will argue provides a picture of the functional data that is both stable to noise, easy to compute, and captures
the important features of each function with no need for horizontal or vertical alignments.
We will then use machine-learning methods to classify the persistence diagrams into behavioral types.

The persistence diagram  is one of the main tools in Topological Data Analysis (TDA),
a new and developing field \cite{Edelsbrunner2000,FrosLand99}
which adapts methods from algebraic topology to find structure 
in complex datasets.
There have already been many promising applications of TDA to (for example) gene expression \cite{Dequeant2008}, signal analysis \cite{perea2013}, and orthodontia \cite{Gamble2010}.

PDs provide a robust and low-dimensional picture of some of the multi-scale topological and geometric 
information carried by a point cloud or a function on some space.
A rigorous explanation of the most general type of PD requires a background in algebraic topology.
Fortunately, the PDs we use in this paper are from a much simpler context, representing the evolution
of connected components for the threshold or sub-level sets of functions $f: [a,b] \to \mathbb{R}$, 
and can be understood with almost no algebraic topology knowledge.
For a survey on PDs in general, see \cite{Edelsbrunner2010}. The requisite background
on algebraic topology can be found in (for example) \cite{Munkres1993}.

Once a PD has been computed, the obvious question is how to interpret it.
In most applications to date, the answer has been rather ad-hoc.
In this paper, we propose what we hope will be a widely-applied general machine-learning method for interpreting 
PD datasets in a statistical context. See the related paper \cite{Learning2014} for more details on this method.

\subsection{Outline.}

The structure of this paper is as follows.
In the next section, we describe a MHT tracker that is particularly useful in tracking applications.
We then demonstrate that enriching this particular MHT tracker with topological features motivated by behavioral characteristics greatly improves the efficacy of the tracker.

In Section \ref{sec:behavior}, we introduce the simple behavior functions associated to each tracklet.
An elementary introduction of persistence diagrams, in the context of these functions, comes in Section \ref{sec:PD}.
The transformation of these diagrams into features suitable for machine learning is described in Section \ref{sec:learning},
which also contains the experiments we ran on simulated data to demonstrate that the learned topological features
do succeed in picking out behavior types.
The technical details behind the incorporation of the topology into the MHT tracker is given in Section \ref{sec:IT}, where
we demonstrate via example that the new tracker connects-the-dots in a situation where the old tracker remains confused.

%
%
%
%

\section{The Tracker}
\label{sec:tracker}

This paper demonstrates the utility of TDA methods by integrating them with an extant MHT tracker. Later in this paper we shall show
that integration appreciably improves results, but we describe the initial paradigm before integration of the modular TDA component.

\subsection{Pre-Topology Tracker.} 
\begin{figure}[t]
\begin{center}
\includegraphics[scale=0.275]{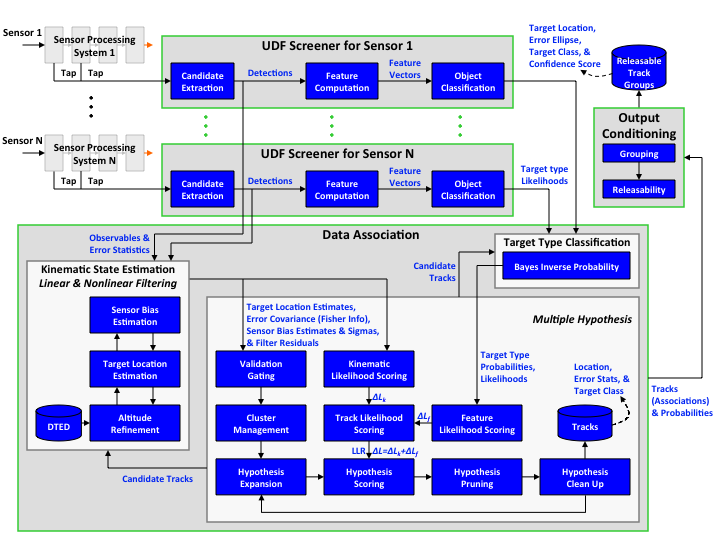}
\caption{Upstream Data Fusion (UDF) MHT tracker}
\label{fig20}
\end{center}
\end{figure}

The core framework of our approach is an MHT tracker which has been used for field-scale applications in areas including mobile missile detection systems, maritime ship tracking, and space situational awareness. This tracker (shown in Figure \ref{fig20} and described in \cite{Newman2013}) was developed for Upstream Data Fusion (UDF) to recover information in upstream data that would otherwise be lost using traditional downstream fusion methods. 
The sensor systems feeding the tracker are shown in the upper left part of the figure with upstream 
data taps indicated. 

The first step in processing is to screen the data to produce observables for kinematic state 
(e.g., position, velocity and acceleration states) estimation and target type features for target type estimation. 
The kinematic state estimation is accomplished with a Square Root Information Filter (SRIF) that uses 
Gauss Newton methods for iterative nonlinear filtering, compensates for sensor system biases, 
and for ground targets uses Digital Terrain Elevation Data (DTED). 
The SRIF and the target type estimation are both Bayes inverse probability solutions.

Data association is also provided by a Bayes inverse probability solution that uses the other inverse 
probability solutions as part of its solution. 
Data association for tracking can be viewed as nested inverse probability problems where the 
kinematic estimation and type estimation are nested inside data association as shown in Figure \ref{fig20}. 
We use this approach to incorporate behavior estimation in MHT by adding another inverse 
problem for target behavior estimation to the nest in a later section. 

The multiple hypothesis block of Figure \ref{fig20} expands multi-track hypotheses, scores them with the 
LLR and prunes the hypotheses down to a reasonable number to carry forward. 
Obviously incorrect data associations are eliminated in a gating process if a track has a statistically 
excessive residual fit error. 
Also, data are clustered into disjoint regions of confusion in a divide and conquer approach.

The basis for the LLR can be found in Reid's fundamental inverse probability expression in Equation (\ref{eq:reid}):

\begin{figure*}[!t]
\begin{equation}
 \label{eq:reid}
\begin{split}
p\left(\Omega_i^k\vert Z^k\right)&= p\left(\Psi_h,\Omega_g^{k-1}\vert Z(k),Z^{k-1}\right)\\
&= \frac{p\left(Z(k)\vert\Psi_h,\Omega_g^{k-1},Z^{k-1}\right) p\left(\Psi_h\vert\Omega_g^{k-1},Z^{k-1}\right) p\left(\Omega_g^{k-1}\vert Z^{k-1}\right)}{p\left(Z(k)\vert Z^{k-1}\right)}
\end{split}
\end{equation}
\end{figure*}

where $\Omega_i^{k}$ is the data association hypothesis, $\Psi_h$ is the current scan data association hypothesis, $\Omega_g^{k-1}$ is the hypotheses through $k-1$ scans, and $Z(k)=\{z_1(k),z_2(k),\ldots,z_{m_k}(k)\}$ is the set of measurements on the current scan at time $t_k$.
 
The solution recursively processes a scan of data to update the probabilities of data association 
hypotheses. 
A scan of data satisfies the scan constraint that, for each track in the hypothesis, there is at most 
one measurement in the scan and, for each measurement in the scan, there is at most one track 
in the hypothesis. Typical imagery data and Ground Moving Target Indicator (GMTI) data satisfy 
the scan constraint. 
Typical signal data do not satisfy the scan constraint so the recursion is done for one measurement 
at a time. 
The left-hand side of the equation is the probability of the multiple track data association hypothesis
 given the data  through time $k$. 
 The first two terms in the numerator of the right-hand side of the equation are the conditional 
 Bayesian likelihood terms. 
 The left term of the two is the conditional likelihood function that is the probability density of the 
 new data given the parent and child data association hypotheses and the previous data. 
 The right term of the two is the Bayesian term that is the probability of the child given the parent and 
 previous data. 
 Since the current data does not enter this term, the probability of how the current data associates 
 is based entirely upon prior information. 
 Thus, it is a Bayesian probability computed from prior quantities such as the probability of detection 
 and false alarm rate of the sensor system.
It is easy to see that the highest probability hypothesis is the one with the highest product of 
conditional Bayesian likelihood terms over time. 
Taking the natural log of this and normalizing it produces the LLR scoring term for association 
hypotheses described by Bar Shalom, et al. in \cite{BarShalom2007}. 
Assuming data in the scan are statistically independent of each other and that the kinematic 
space data and target type feature space data are independent of each other, the likelihood 
term can be expressed as
\begin{equation}
\label{eq:likelihood}
\begin{split}
p\left(Z(k)\vert\Psi_h,\Omega_g^{k-1},Z^{k-1}\right)=&\displaystyle\prod_{j=1}^{m_k} p\left(z_j(k)\vert\Psi_h,\Omega_g^{k-1},Z^{k-1}\right)\\
p\left(z_j(k)\vert\Psi_h,\Omega_g^{k-1},Z^{k-1}\right)=&\: p\left(z_{ks,j}(k)\vert\Psi_h,\Omega_g^{k-1},Z^{k-1}\right) \\
&\times p\left(z_{fs,j}(k)\vert\Psi_h,\Omega_g^{k-1},Z^{k-1}\right)
\end{split}
\end{equation}

The natural log of the kinematic space term can be computed using SRIF quantities as described in 
\cite{Bierman1990} to be the following
\begin{eqnarray}
\label{eq:srif}
-\frac{1}{2}\left(e_k^te_k+2\frac{\log\det R_{k/ k}}{\log\det R_{k/k-1}}\right)
\end{eqnarray}
The $e_k$ term is the SRIF residual data fit error and the $R_{\cdot/\cdot}$ term is the square root of the information 
matrix where the information matrix is the inverse of the covariance of the state estimate error.

For target type feature data, suppressing the association hypothesis, assume a finite number  
of exhaustive, mutually exclusive discrete target types and sum a joint distribution over the target 
types to obtain the likelihood as shown in the first expression in Equation (\ref{eq:joint}). 
\begin{figure*}[!t]
\begin{equation}
\label{eq:joint}
\begin{split}
p\left(z_{fs,k}\vert\Psi_h,\Omega_g^{k-1},Z_{fs}^{k-1}\right)=&\left\{\begin{array}{cl}
																		p(z_{fs,k}\vert FT)&\text{for false target}\\
																		\displaystyle \sum_{i=1}^{N_T} p(z_{fs,k},T_i\vert Z_{fs}^{k-1})& \text{for detected target}
																		\end{array}\right.\\
\sum_{i=1}^{N_T} p(z_{fs,k},T_i\vert Z_{fs}^{k-1})=& \sum_{i=1}^{N_T} p(z_{fs,k}\vert T_i,Z_{fs}^{k-1}) p(T_i\vert Z_{fs}^{k-1})\\
p(T_i\vert Z_{fs}^k)=&\frac{p(z_{fs,k}\vert T_i, Z_{fs}^{k-1}) p(T_i\vert Z_{fs}^{k-1})}{p(z_{fs,k}\vert Z_{fs}^{k-1})}	
\end{split}		
\end{equation}
\end{figure*}
But the joint distribution can be expressed using Bayes rule in terms of the conditional likelihood 
conditioned on the target type and the probability of the target type given past data as shown in 
the second expression in Equation (\ref{eq:joint}).
Further, the probability of the target type can be provided by the inverse probability solution shown 
in the last expression in Equation (\ref{eq:joint}). 
This last expression is a recursive Bayes inverse probability solution for target type classification. 
The key quantity needed in the above is the conditional likelihood that can be determined using 
machine learning methods as described in \cite{Newman2013}.

The target type feature data can actually be target type classification computed using only the 
current data. 
Then the conditional likelihood above is the confusion matrix of the classifier as described in 
\cite{BarShalom2005}. 
An upstream solution is to use target type features directly, rather than actually computing a target type 
classification from the current data in a track. 
The recursive Bayes inverse probability solution provides a target type classification using all the 
feature data in the track.

It can be shown that the likelihoods provided by Equation (\ref{eq:srif}) and Equation (\ref{eq:joint}) are mathematically equivalent by deriving Equation (\ref{eq:srif}) in terms of the integral over the continuous kinematic state space of the 
product of the conditional likelihood with the probability density of the kinematic state conditioned 
on the past data. This reveals the nested inverse probability problem nature of data association. 
This nesting is continued later in this paper to include target behavior estimation.

\section{Behavior Functions}
\label{sec:behavior}

We now introduce a few simple functions that will be used to characterize aspects of driver behavior.
The functions used in the work presented here are based on speed, acceleration, and turning radius,
but the general framework could certainly incorporate different function types.

\subsection{Functional profiles.}




Suppose we have a tracklet under consideration which has coordinates
\[
 \{(x(t_i),y(t_i),z(t_i): 0 \leq i \leq n\},
 \]
where $n$ is the number of pixels in the tracklet.
Let $v(t_i)$ be the resulting velocity vector, computed for example as a $2$-step
average:
\[
v(t_i) =(\Delta x_i,\Delta y_i,\Delta z_i)
\]
with
\[
\Delta x_i = \frac{x_{i}-x_{i-1}}{t_i - t_{i-1}},\: \Delta y_i = \frac{y_{i}-y_{i-1}}{t_i - t_{i-1}},\text{ and } \Delta z_i = \frac{z_{i}-z_{i-1}}{t_i - t_{i-1}}.
\]
From the $v(t_i)$ we compute other quantities as follows.

\subsection{Simple Function Choices}

The first function we use is the speed curve of each vehicle:
\[
s(t_i) = \Vert v(t_i)\Vert.
\]
We will also consider the acceleration
\[
a(t_i) = \Vert\dot{v}(t_i)\Vert
\]
estimated similarly to $v$,
and turning, which is a variant of the angular velocity 
\[
T(t_i) = \frac{\Vert v(t_i)\Vert + \Vert v(t_{i+1})\Vert}{2} sin(\theta_i)
\]
where $\theta_i$ is the angle between $v(t_i)$ and $v(t_{i+1})$.

Our hypothesis is that both the speed and the acceleration functions should display strong variation
for a more aggressive driver.
The turning function measures how fast agents go when turning.
When an agent turns, $\sin(\theta)$  is non-zero, and a faster turn will 
mean that the norms of the velocity vectors are larger.
A car going fast in a straight line won't register, although a ``fast lane-switcher'' 
may have small $\sin$ values but very large velocity so may register, but they
will give a different profile.
We look for a large peak for racers, and a smaller peak for slow drivers.

\subsection{A poor choice: transient critical values.}

Since we are interested in {\em practical and computable} invariants of track motion, one
might ask why we could not simply use the maximum speed or turning values computed
from vehicle velocities.
It is easy to see that this is not a particularly good invariant of behavior, and we illustrate
this fact with a simple story.

\begin{figure}[ht]
\begin{minipage}[b]{0.45\linewidth}
\centering
\includegraphics[width=\textwidth,scale=0.2]{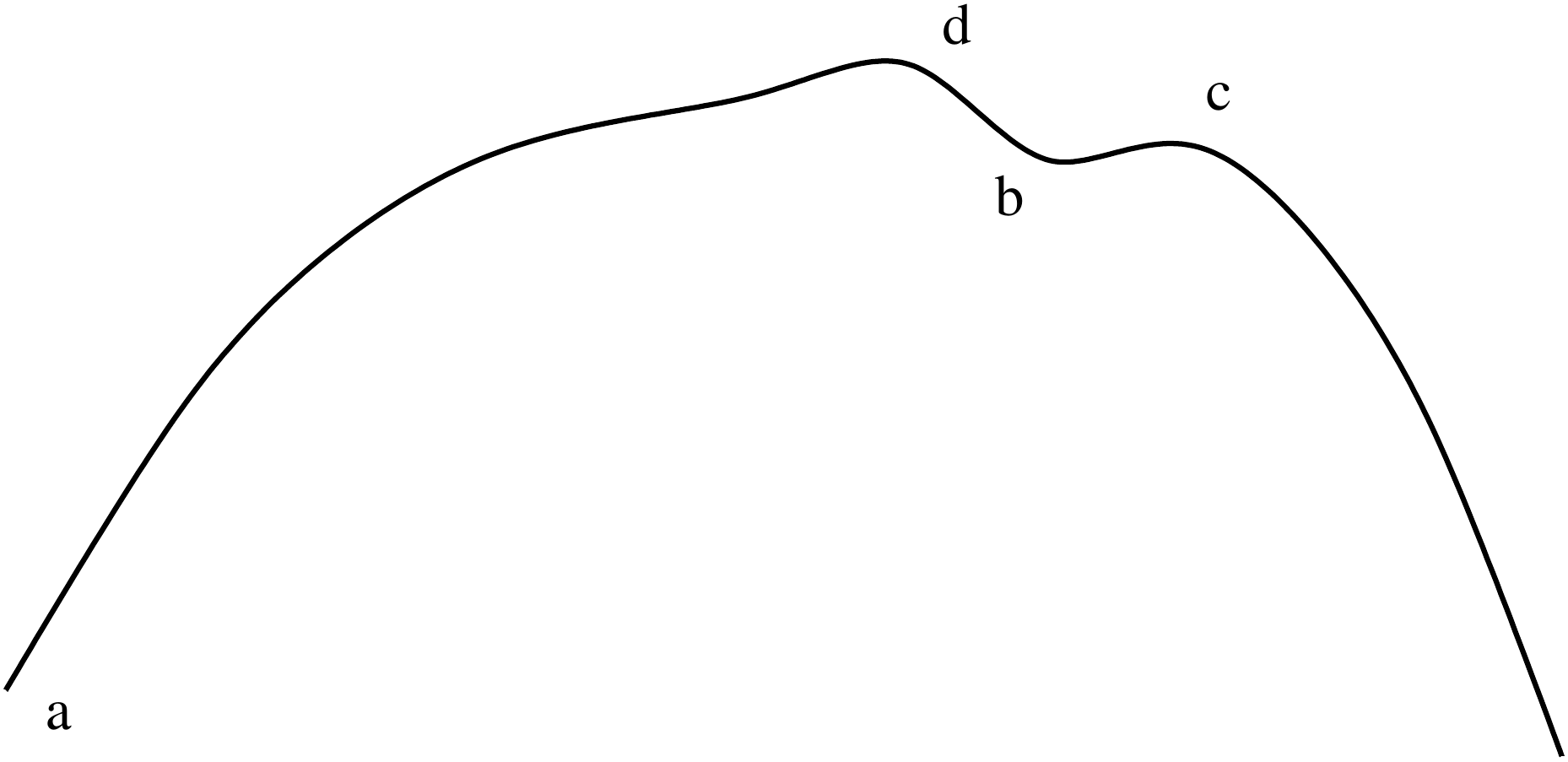}
\end{minipage}
\hspace{0.5cm}
\begin{minipage}[b]{0.45\linewidth}
\centering
\includegraphics[width=\textwidth,scale=0.2]{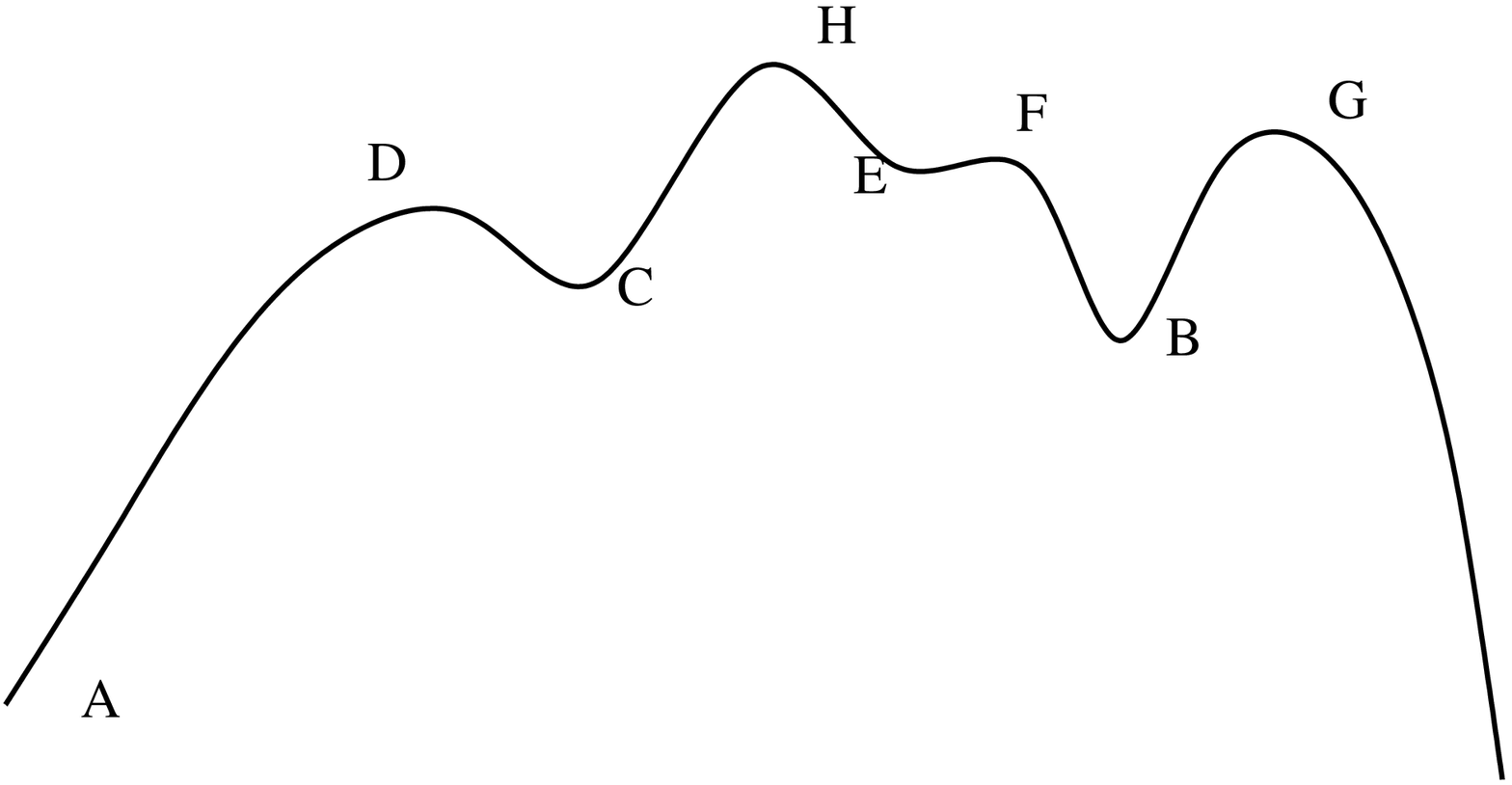}
\end{minipage}
\caption{Possible speed profiles for a non-agressive driver (left), and an aggressive one (right).
The letter labels indicate the critical values.
 }
\label{fig:speeds}
\end{figure}

Consider the two speed functions shown in Figure \ref{fig:speeds}.
We imagine two vehicles which our camera captures entering and then leaving a freeway with a posted
speed limit of $65$ miles per hour (MPH).
Both vehicles enter the highway, and speed up to $70 MPH$.
Vehicle $A$ speeds up evenly, maintains mostly constant speed near $70$ while on
the freeway, and slows down gradually at the exit ramp.
Vehicle $B$ accelerates more quickly (guns it), speeds up and slows down during
the time on the freeway, perhaps because it passes many cars and tailgates others, and then exits with a sharp deceleration.

The maximum speed may capture very little information to distinguish these two.
But if we look at the shape of the speed curves, we see considerably more variation
for $B$. Furthermore, most of this extra variation happens at high speed values.
The persistence diagram of the speed function, described in the next section, is ideally suited to summarize
all of this information.


\section{Persistence Diagrams}
\label{sec:PD}

In this section, we introduce the $0$-dimensional \emph{persistence diagram} $D_0(f)$,
of a function.
We will not do so in the most general context, choosing instead to restrict to what is most relevant to
this particular application.

\subsection{$0$-dimensional persistence for continuous functions}
We start with a continuous function 
$f: [a,b] \to \mathbb{R}$.
Typically $f$ is given by its sample values at a set of points
$T = \{t_0, \cdots, t_n \} \subset [a,b]$,
$a = t_0 < t_1 < \cdots < t_n = b$, 
and we linearly interpolate to obtain a continuous function.
Let $u_i = f(t_i)$, and assume for simplicity that $u_i \neq u_j$ for all $i \neq j$.
If this is not the case, a small deformation of values breaks
the ties and corrects the situation.
Let $m$ and $M$ be the minimum and maximum (respectively) of $f$ on $[a,b]$. By continuity, $m$ and $M$ occur at points in the set $T$.
 
Define the sub-level set at level $c$ of $f$ to be 
\[
f_c = f^{-1}\left((-\infty,c] \right)= f^{-1}\left([m,c]\right)
\]
By continuity of the function $f$, we know that $f_c$ has the form
\[
f_c =\bigcup_{k=1}^K [a_k,b_k]
\]
where each of the closed intervals $[a_k,b_k]$ are disjoint and satisfy $a_k\leq b_k$. Given such a decomposition of $f_c$, we say that $f_c$ has $K$ \emph{components}, and we track how many components $f_c$ has as the level $c$ goes from $m$ to $M$. A real number $c$ is called a \emph{critical value} of $f$
if  $f_c$ has either more or less components than $f_{c-\epsilon}$
for $\epsilon$ sufficiently small\footnote{Small enough that $u_j \notin [c-\epsilon,c)$
for any $j$.}. For example, the critical values of the functions in Figure \ref{fig:speeds}
are labeled by letters.

If the number of components goes up when we pass the threshold value
$c$, we say that a \emph{birth} happens at $c$.
In our interpolated case, this happens at a point $t_i$
where both $u_{i-1} > c = u_i$ and  $u_{i+1} > c$.
On the right side of Figure \ref{fig:speeds}, the births
happen at $A,B,C,$ and $E$; on the left side, they happen
at $a$ and $b$.

We say that a {\em death} occurs at $c$ if 
$f_c$ has fewer components than $f_{c-\epsilon}$
for $\epsilon$ small.
In our interpolated case, this happens at a point $t_i$
where both
$u_{i-1} < c = u_i$ and  $u_{i+1} < c$.
These happen at $D,F,G,$ and $H$ for the aggressive speed function,
and at $c$ and $d$ for the normal one.

The main idea of persistence is to create a pairing of  the births 
and deaths of $f$.
To do this, we associate to each birth the height 
$u_i$
where the component was born.
Then, when we come to a death, we are merging two
components.
The {\em younger} of the two is the one with the higher
 birth value.
 We then say that the younger one dies at the merger since its component is merged with
one that existed before it was born.

If a component was born at height $u_i$ and dies at
height $u_j$ we record the pair $(u_i,u_j)$ and say
that the persistence of the component is $u_j - u_i$.
Persistence measures how long a component existed.
In the case we are discussing this can be interpreted as
the  size of the shape feature represented by the component.

For the aggressive speed function,
the persistence pairs are $(A,H), (C,D), (E,F),$ and $(B,G)$;
for the normal one, they are $(a,d)$ and $(b,c)$.

\subsection{Diagrams}

\begin{figure}[ht]
\begin{minipage}[b]{0.45\linewidth}
\centering
\includegraphics[width=\textwidth,scale=0.2]{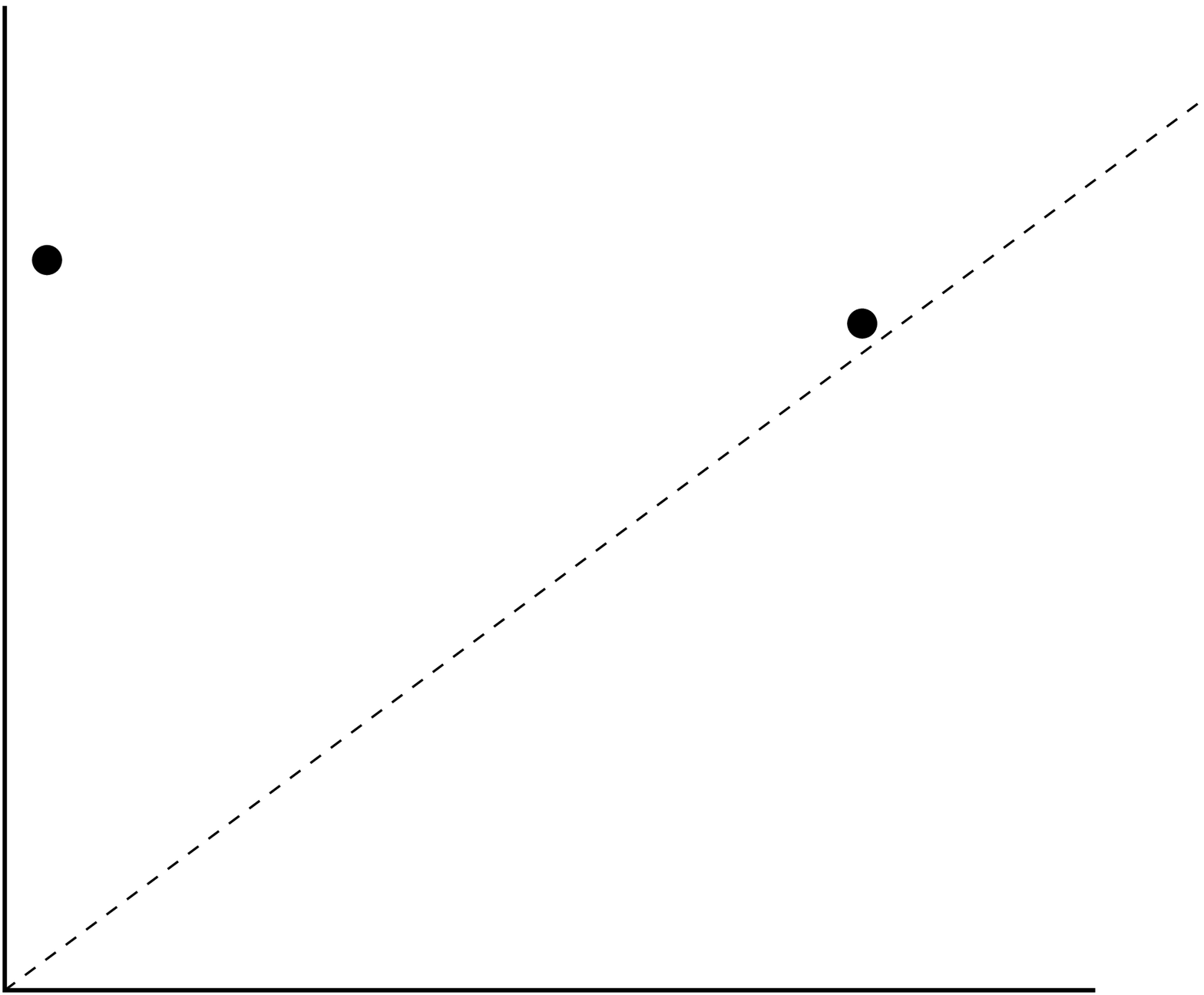}
\end{minipage}
\hspace{0.5cm}
\begin{minipage}[b]{0.45\linewidth}
\centering
\includegraphics[width=\textwidth,scale=0.2]{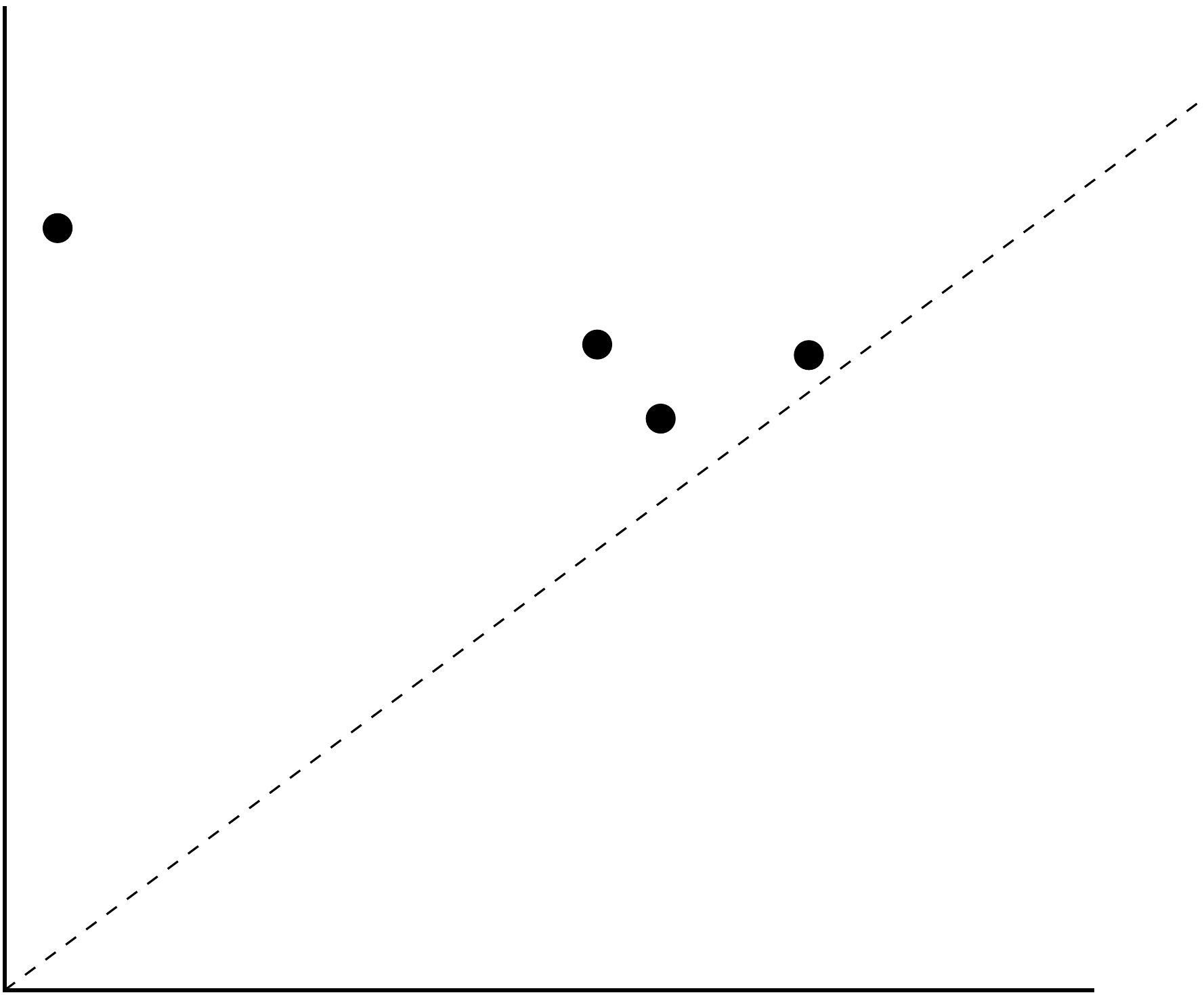}
\end{minipage}
\caption{The persistence diagrams $D_0(f)$ (left) and $D_0(g)$ (right), where
$f$ and $g$ are the speed functions on the left and right, respectively, of
Figure \ref{fig:speeds}.
Note that both diagrams have the same highest-persistence dot, which corresponds
to their identical max speeds. 
The extra dots in the right diagram correspond to the extra variation in $g$.
 }
\label{fig:diagrams}
\end{figure}

All of the above birth, death, and persistence information is summarized
compactly in the \emph{persistence diagram} $D_0(f)$.
This is a multi-set of dots in the plane, containing one
dot $(u_i,u_j)$ for each component born at $u_i$ and killed at $u_j$.
See Figure \ref{fig:diagrams} for the persistence diagrams of
the two speed functions from Figure \ref{fig:speeds}.

We note that these diagrams allow one to hunt for some common functional motifs among a function
population withot any need for pre-alignment in either the horizontal or vertical direction.
For example, the question of whether two functions have a min-max pair of a certain size, independent of
where the critical points occur in the interval domains, can be settled simply by looking at their respective persistence
diagrams.
Of course, this process can be automated, as described in the next section.

\subsection{Stability.}

An important feature of persistence diagrams $D_0(f)$ is that they are robust to small
changes in the input function $f$.

To make this precise, we need a metric on the set of persistence diagrams,
so we now define the \emph{Wassertein} distance $W_p(D_0(f), D_0(g))$ between
two diagrams. First, we fix some $p \in [1, \infty]$.
We then adopt the convention that every diagram contains
a dot of infinite multiplicity, and zero persistence, at every point $(u,u)$
along the major diagonal in the plane.
For each bijection $\phi: D_0(f) \to D_0(g)$,
we define its cost to be 
$$
C_p(\phi) = \left(\sum_{u \in D_0(f)} ||u - \phi(u)||_p \right)^{\frac1p};
$$
note
that such bijections exist even if the two diagrams have different numbers
of off-diagonal dots, as we can always match extra dots to the diagonal dots 
in the other diagram.
Finally, $W_p(D_0(f), D_0(g)$ is defined to be the minimum possible
cost $C_p(\phi)$, as $\phi$ ranges over all possible bijections between the diagrams.

We remark that $W_1$ is often called the \emph{earth-mover distance}, while
$W_{\infty}$ is the \emph{bottleneck distance}.
In addition, any of the distances $W_p$ can be computed via a max-flow-min-cut algorithm.
\begin{figure}[h]
 \begin{center}
  \includegraphics[scale=0.3]{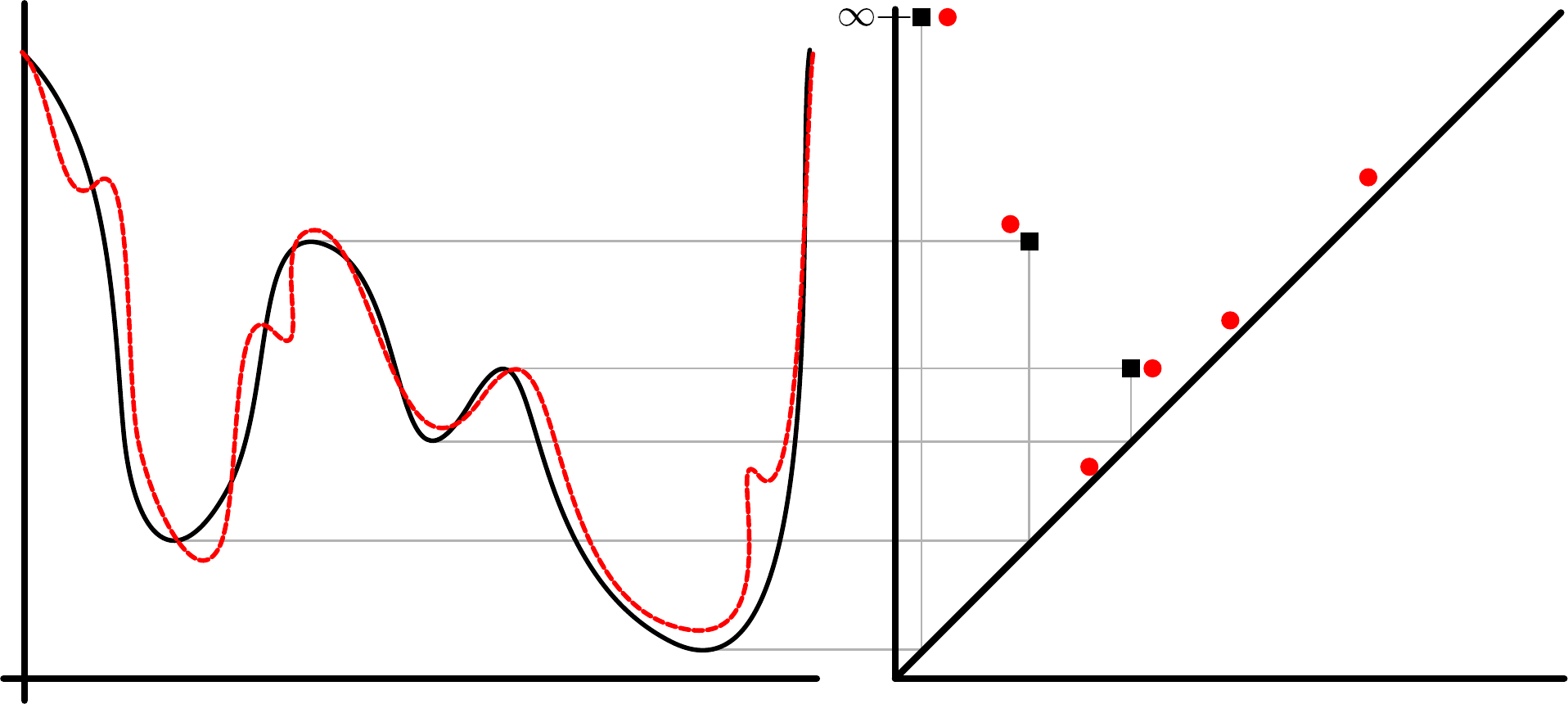}
 \end{center}
\caption{Demonstration of stability. The graphs and persistence diagrams for a function
and its noisy version
are shown in black and red, respectively.
The optimal bijection $\phi$ matches the high-persistence dots with each other, and the low-persistence
red dots with the diagonal.}
\label{fig:noise}
\end{figure}

There are then several different stability theorems \cite{CohenSteiner2007,Chazal2009b,Cohen-Steiner2010} which state
that, under very mild conditions, $W_p(D_0(f), D_0(g)) \leq C ||f - g||_{\infty}$.
In other words, a small variation in the input function will not cause
a large change in the output diagram. Figure \ref{fig:noise} illustrates this phenomenon.

\subsection{Augmented Morse filtration.}

In practice, we are given the finite set of values $\{f_i=f(t_i)\}_{i=0}^n$ and we compute the diagram using the discrete Morse filtration procedure. However, this algorithm discards a large amount of information concerning the values that the function attains. To avoid this waste, we augment the discrete Morse filtration procedure as described in Algorithm \ref{algo:MF}. This merely appends additional values to the diagonal of the the resulting persistence diagram. It should be noted that the stability results discussed in the last subsection also apply to these augmented diagrams.

The time complexity of both of these algorithms is dominated by the sorting procedure, and so these algorithms are ${O}(n\log n)$. The space complexity of these algorithms is ${O}(n)$.

\begin{algorithm}[h]
\caption{The (augmented) discrete Morse filtration algorithm.} \label{algo:MF}
\begin{algorithmic}
\STATE Input $\{(i,f_i)\}_{i=0}^n$
\STATE Initialize diagram $D_0\gets \emptyset$, and class indicators $\{c_i\}_{i=0}^n\gets \{-1\}_{i=0}^n$
\STATE Sort $\{(i_k, f_{i_k})\}_{k=0}^n$ so that $\{f_{i_k}\}_{k=1}^n$ is in ascending order
\FORALL{ $k=0,\ldots,n$ }
	\STATE ${N}\gets \emptyset$
	\IF{ $i_k > 0$ \AND $c_{i_k-1}\not = -1$}
		\STATE ${N}\gets {N}\cup\{i_k - 1\}$
	\ENDIF
	\IF{ $i_k < n$ \AND $c_{i_k+1}\not = -1$}
		\STATE ${N}\gets {N}\cup\{i_k - 1\}$
	\ENDIF
	\IF{ ${N}=\emptyset$ }
		\STATE $c_{i_k}\gets k$
	\ENDIF
	\IF{ ${N}=\{ j \}$ }
		\STATE $c_{i_k}\gets c_j$
		\STATE $D_0\gets D_0\cup\{(f_{i_k},f_{i_k})\}$ {\bf IF AUGMENTED}
	\ENDIF
	\IF{ ${N}=\{i_k-1,i_k+1\}$ }
		\STATE $c_{\text{oldest}} \gets \min(c_{i_k-1}, c_{i_k+1})$
		\STATE $c_{\text{youngest}} \gets \max(c_{i_k-1}, c_{i_k+1})$
		\FORALL{ $c_i = c_{\text{youngest}}$}
			\STATE $c_i \gets c_{\text{oldest}}$
		\ENDFOR
		\STATE $D_0\gets D_0\cup\{(f_{c_{\text{youngest}}}, f_{i_k}) \}$
		\STATE $D_0\gets D_0\cup\{(f_{i_k},f_{i_k})\}$ {\bf IF AUGMENTED}
	\ENDIF
\ENDFOR
\RETURN $D_0$
\end{algorithmic}
\end{algorithm}

\section{Learning Behavior from Diagrams}
\label{sec:learning}

As we have seen, diagrams can encode the repetitive structure of behaviors. In order to exploit this encoding, we would like to project persistence diagrams into a Euclidean feature space so that distinct classes are rendered linearly separable. Ideally, the encoding of diagrams in this feature space should be sparse (i.e. the entries of the resulting feature vector should consist mostly of zeros). This reflects the desire that most objects should be sufficiently described using only a few of the atomic features represented by standard orthogonal directions in the feature space.

\subsection{Binning and feature vectors.}

\begin{figure}[h]
\begin{center}
\includegraphics[scale=0.4]{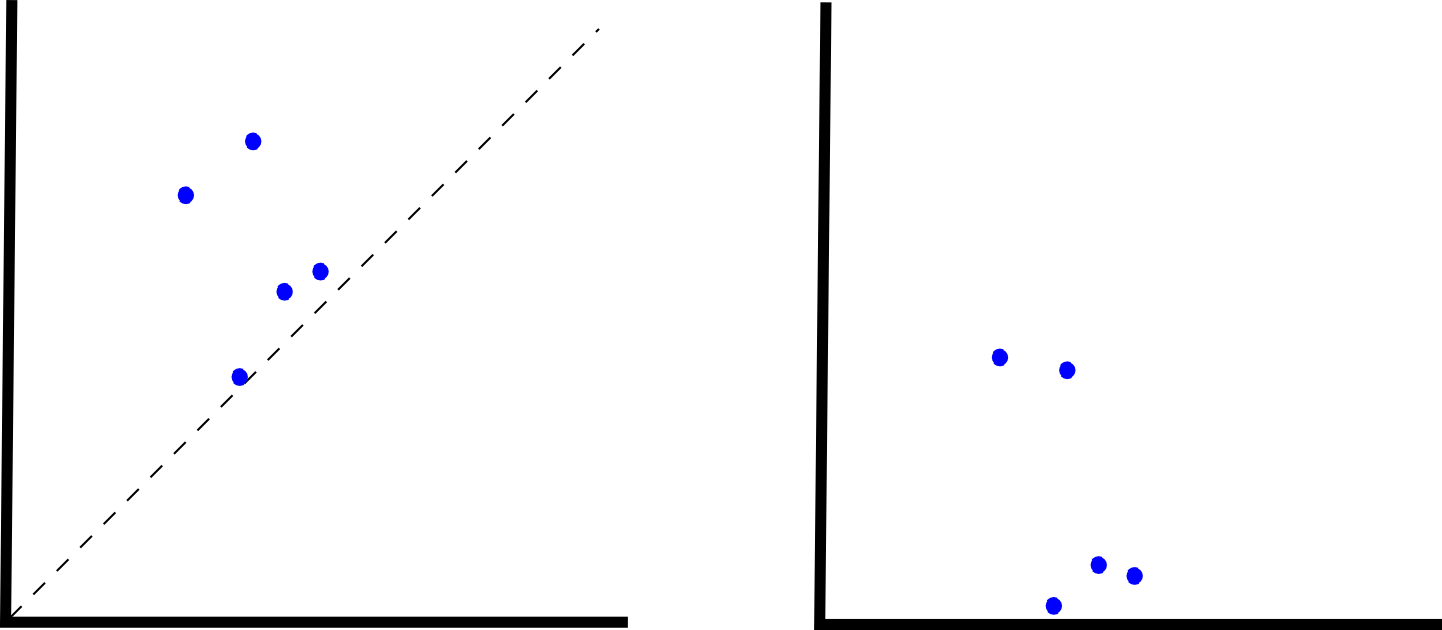}
\end{center}
\caption{Persistence diagram transformation.}
\label{fig:transform}
\end{figure}

For signals exhibiting repetitive structure, the visualization of persistence diagrams clearly indicates the spatial isolation of birth-death pairs. This suggests that we may employ regular tilings of the plane to ``bin'' the persistence diagrams, thereby obtaining sparsely-structured feature vectors. Moreover, when classes of signals exhibit distinct critical point behavior, projecting persistence diagrams in such a way induces linearly separable classes. 
	We now describe the details of this projection. Given a persistence diagram $D=\{(\alpha_i,\beta_i)\}_{i=1}^N$, we first map to $\widetilde{D}=\{(\alpha_i,\beta_i-\alpha_i)\}_{i=1}^N$ as depicted in Figure \ref{fig:transform}. This skew-transformation just aligns the persistence diagram so that rectangular tilings of the upper-half plane are compatible with the natural constraints of the diagram. 

\begin{figure}[h]
\begin{center}
\includegraphics[scale=0.8]{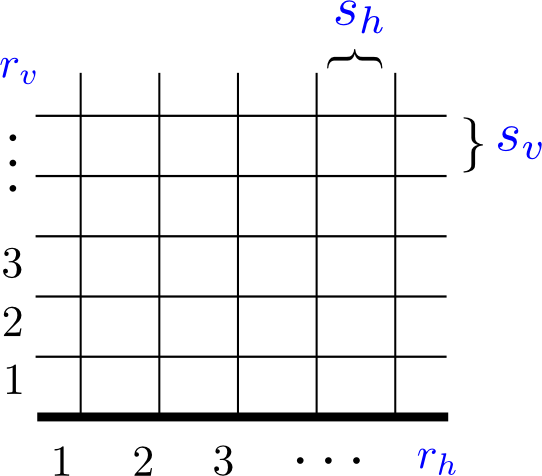}
\end{center}
\caption{Partitioning the upper-half plane for binning.}
\label{fig:binned}
\end{figure}

Now, we fix a parameter vector 
\[
\omega\in \{ (r_v, r_h,\alpha_0,\beta_0,\beta_1) : r_v, r_h\in \{4,5,\ldots\},\: \alpha_1>\alpha_0 > 0,\:\beta > 0\}.
\]
The natural numbers $r_v$ and $r_h$ indicate the vertical and horizontal resolution of the ``binned" diagram. The $\alpha$ values indicate the range of ``birth" values that are under scrutiny, and the $\beta$ value indicates the range of ``lifetime" values that are under scrutiny. For simplicity, we define the auxiliary parameters
\[
s_v = (r_v - 1) / \beta\text{ and } s_h = (r_h - 2) / (\alpha_1 - \alpha_0)
\]
to indicate the height and width of our rectangular bins. We also define the partition $\mathcal{I}=\{I_i\}_{i=1}^{r_v}$ of $(0,\infty)$ by
\[
I_i =\left\{\begin{array}{ll}
						(\beta,\infty) & i = 1\\
						( (r_v - i) s_v, (r_v - i +1)s_v] & i > 1
						\end{array}\right.
\]
and the partition $\mathcal{J} = \{J_j\}_{j=1}^{r_h}$ of ${R}$ by
\[
J_j =\left\{\begin{array}{ll}
						(-\infty,\alpha_0] & j = 1\\
						 ((j-1)s_h, j s_h]& 1 < j < r_h\\
						 (\alpha_1,\infty) & j = r_h
						\end{array}\right.
\]
These partitions will induce a partition of the upper-half plane as depicted in Figure \ref{fig:binned}.

We now define the matrix $\text{bin}_\omega(\widetilde{D})\in \mathbb{R}^{r_v\times r_h}$ (the space of $r_v$ by $r_h$ matrices) by setting
\[
\left(\text{bin}_\omega(\widetilde{D})\right)_{i,j} = \left\vert \left(I_i\times J_j\right) \cap \widetilde{D}\right\vert
\]
for all pairs $(i,j)$ with $1\leq i\leq r_v$ and $1\leq j\leq r_h$. From this definition, it is clear that the partition elements $I_1$, $J_1$, and $J_{r_h}$ are ``overflow" regions that ensure the invariance of total counts under projection.
 
The savvy reader may note that this tiling of the plane for the purposes of constructing histograms is somewhat arbitrary, and could itself be learned or optimized. One potential strategy for bin selection would optimize over a fixed number of Voronoi regions. However, a reasonably fine resolution generally captures these regions with sufficient clarity, while also remaining compatible with simple visualization techniques (bitmaps).

\subsection{Models for binned diagrams.}

Here, we introduce several models for analyzing binned diagrams. \\

\noindent {\it Logisitic regression}: In this scenario, we are simply interested in determining how to separate the binned diagrams of distinct classes using hyperplane arrangements. For each class $c$, we have parameters ${\bf \theta}_c$ and $b_c$. The probability of assigning an object with feature vector ${\bf x}$ to class $c$ is then given by
\[
p(c\vert {\bf x}) = \frac{\exp\{-\langle {\bf \theta}_c, {\bf x}\rangle -b_c\}}{\sum_{c^\prime} \exp\{-\langle {\bf \theta}_{c^\prime}, {\bf x}\rangle -b_{c^\prime}\}}.
\]
Training these parameters is done via the stochastic gradient descent procedure described in \cite{zhang2004solving}.\\

\noindent {\it Poisson process}: To more accurately reflect the fact that our feature vectors have integral entries, we may model the counts as arising from several independent Poisson-distributed random variables. Because the number of total counts of such a model can be variable, this model encodes a certain amount of domain invariance. The likelihood of a particular feature vector given a particular class under the Poisson model is given by
\[
p({\bf x}\vert c) = \prod p_{\text{Poisson}}(x_i\vert \lambda_{c,i}).
\]
where $p_{\text{Poisson}}(x\vert\lambda) = \frac{\lambda^{-x}}{x !} e^{-\lambda}$. Classification is then performed using maximum likelihood estimation. 
	Fitting the Poisson rates for each class given a set of labeled data is accomplished using the maximum likelihood estimate for ${\bf \lambda}_c$. That is, given examples $X_1,\ldots, X_N$ from class $c$, we set
\[
{\bf \lambda} = \frac{1}{N}\sum_{i=1}^N X_i.
\] 

\noindent {\it Multinomial}: To capture the relative domain invariance of persistence diagrams more precisely, we may model the counts as a draw from a multinomial distribution where the number of classes is an input parameter. In this case, the likelihoods are given by
\[
p({\bf x}\vert c) = \binom{\vert {\bf x}\vert}{x_1,\ldots,x_K}\prod \theta_{c,i}^{x_i}
\]
for parameter vectors ${\bf \theta_c}$ satisfying $\theta_{c,i}\geq 0$ for all $i$ and $\vert{\bf \theta}_c\vert = \sum \theta_{c,i}=1$.

Fitting these parameters again proceeds using the maximum likelihood estimate, whence
\[
{\bf \theta}_c = \frac{1}{\sum_{i=1}^N \vert X_i\vert} \sum_{i=1}^N X_i
\]
where $X_1,\ldots, X_N$ are the examples from class $c$. 

\subsection{Experimental results.}

The figures referenced in this section can be found in the Appendix. For our experiments, we examine a data set generated using the Simulated Urban Mobility (SUMO \cite{SUMO2012}, http://sumo.sourceforge.net)  platform mainly 
developed by the Institute of Transportation Systems at the German Aerospace Center 

This data set consists of 1000 different vehicular paths, with each path consisting of around 360 triples of positional coordinates indexed by time. Half of the vehicular paths exhibit ``normal" behavior, while the other half exhibit ``aggressive" behavior typified by speeding and accelerations. Typical speed profiles for the classes are illustrated in Figures \ref{fig:norm_spds} and \ref{fig:agg_spds}, the resulting augmented persistence diagrams are illustrated in Figures \ref{fig:norm_diagrams} and \ref{fig:agg_diagrams}, and the corresponding binned diagrams are presented in Figures \ref{fig:norm_bin_diagrams} and \ref{fig:agg_bin_diagrams}.

First, we examined the behavior of our classifiers for the speed (Figure \ref{fig:spd_err}), acceleration (Figure \ref{fig:acc_err}), and turning (Figure \ref{fig:trn_err}) functions described in Section \ref{sec:behavior}. In addition, we considered using all these features combined in Figure \ref{fig:cmb_err}. From the training results, we see that the speed function provides the best test error, and the combined features only marginally improves the classification error. 
Our very best result is a less than $2$ percent test error. 

To emphasize the invariance properties of the persistence diagrams, we consider situations where the length of the input windows (that is, the number of speed values used to compute the different diagrams) varies over both training and testing. We consider window lengths of size 5, 10, 15, 20, 25, and 30 for both training and testing. The clear takeaway is that training can be done on relatively small sizes, but classification error always suffers in the low sampling regime. The estimated standard deviations of these errors over the 25 runs are all less than $10^{-28}$, indicating the robustness of the classification results with respect to the different input orders. The results are depicted in Figures \ref{fig:LRerr}, \ref{fig:Perr}, and \ref{fig:Merr}.

Finally, we illustrate the parameters learned by the logistic regression, Poisson process, and multinomial model in Figures \ref{fig:lrcoeff}, \ref{fig:prates}, and \ref{fig:mprobs} respectively. These figures indicate that the important bins for each class are generally the same for each model.

\section{The Improved Tracker}
\label{sec:IT}

The above experiments demonstrate that our topological behavior features do well in detecting agent behavior.
In this section, we describe how we incorporate these features into an improved MHT.

\begin{figure}[h]
\begin{center}
\includegraphics[scale=0.275]{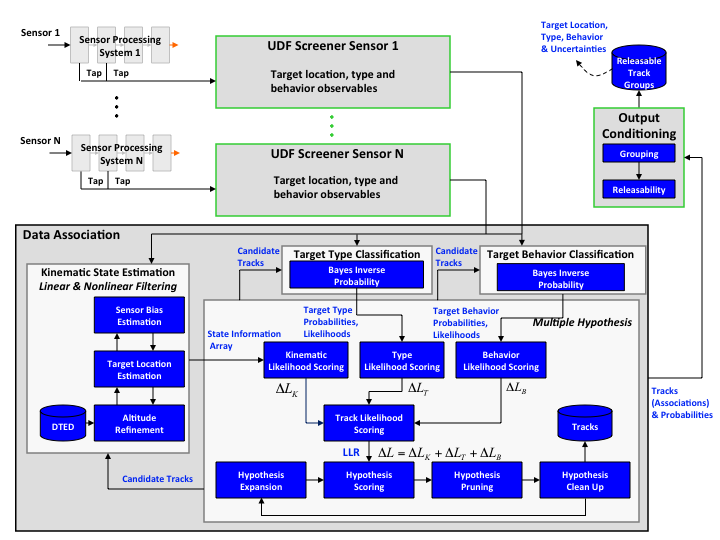}
\caption{UDF MHT tracker using topological features}
\label{fig21}
\end{center}
\end{figure}

\paragraph{Unified Theory.}
A unified tracking theory is provided by combining the UDF MHT tracker with behavior classification using 
topological features. 
The block diagram in Figure \ref{fig21} shows how the UDF MHT tracker of Figure \ref{fig20} is generalized to include 
behavior classification. 

The screener blocks and the multiple hypothesis block have been simplified to focus attention on the 
changes made to incorporate behavior classification. 
An additional inverse problem for target behavior classification has been included in the nested 
inverse problem formulation for data association. 
Changes in the LLR score for a track due to the current kinematic, type, 
and behavior data, respectively, are summed to obtain the total track score. 
This means that data are associated based on the consistency of target kinematic, type, and behavior data. 

The equations that incorporate the target topological features are analogous to Equations (\ref{eq:likelihood}) and (\ref{eq:joint})
for target type features. 
The target type feature space data are replaced by target behavior topological feature space data, 
and target types   are replaced with target behaviors. 
As mentioned previously, the recursive updating is a multiple rate process where the behavior 
LLR updates are done at a slower rate than the kinematic and type updates. 
The topological features are based on behavior functions such as speed, acceleration, and turning. 
The behavior topological features are computed over an interval of time sufficient for the behavior 
to be observed. Then an LLR update is computed.

It is assumed in Equation (\ref{eq:likelihood}) that the feature data are statistically independent of the other data. 
This is also assumed for topological feature data. 
Although speed, acceleration, and turning are computed from kinematic data, 
it is not expected that 
the overall topological shape features will have much correlation with SRIF residuals. 
Also, the behavior functions considered here are computed from SRIF kinematic state estimates. 
Consequently, the estimate errors need to be taken into account when performing machine 
learning to determine conditional likelihood functions.

\paragraph{Tracker Numerical Examples.}
Tracking using topological features provides tracker designers a powerful new way to achieve 
successful tracking by incorporating target behavior understanding. 
We performed numerical experiments for difficult urban tracking with airborne EO sensing 
that is described below. 
As in our previous numerical examples for the classification task, data sets were generated using SUMO. 
A sensor measurement model and error model for an airborne EO sensor were used to produce 
simulated data.

\begin{figure}[!h]
\begin{center}
\includegraphics[scale=0.5]{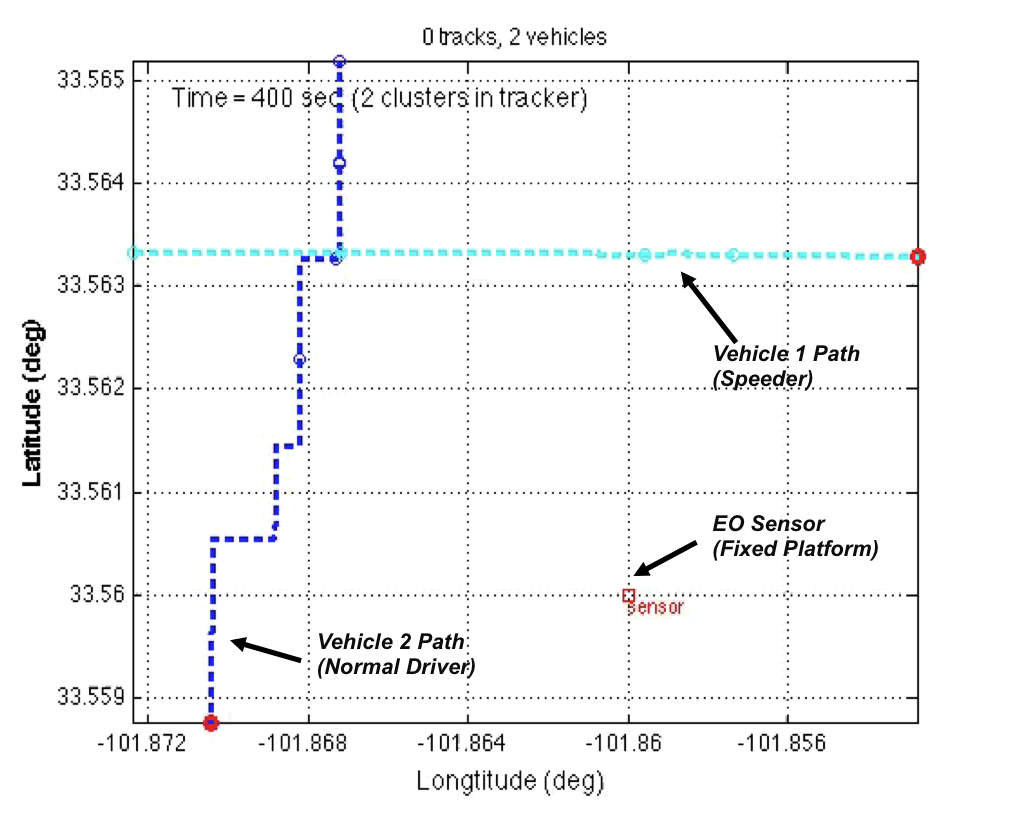}
\caption{Scenario with a speeder and a normal driver observed by airborne EO.}
\label{fig22}
\end{center}
\end{figure}

A simple scenario is shown in Figure \ref{fig22} with two vehicles where one has a driver who is a 
speeder and one who is a normal driver. 
Sensor detection degrades as vehicles slow and stop, for instance, at an intersection. 
Tracking results will be shown for one example with and without topological features that 
focus on the difficulty of successfully tracking through an intersection. 
If a vehicle of interest has been identified but cannot be tracked through an intersection, then it will be lost. 
Further, Monte Carlo results will be shown comparing performance in tracking through the intersection 
with and without topological features with varying levels of angular measurement errors.

\begin{figure}[h]
\begin{subfigure}{0.475\textwidth}
\centering
\includegraphics[scale=0.5]{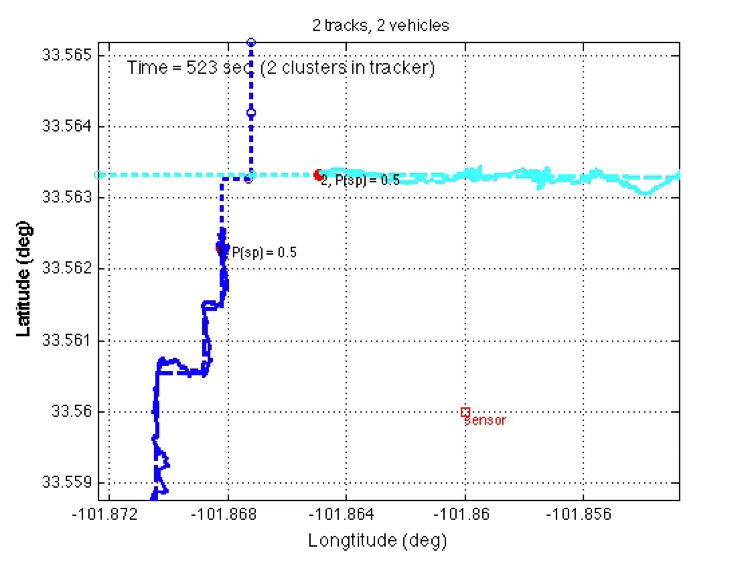}
\end{subfigure} %
\begin{subfigure}{0.475\textwidth}
\centering
\includegraphics[scale=0.5]{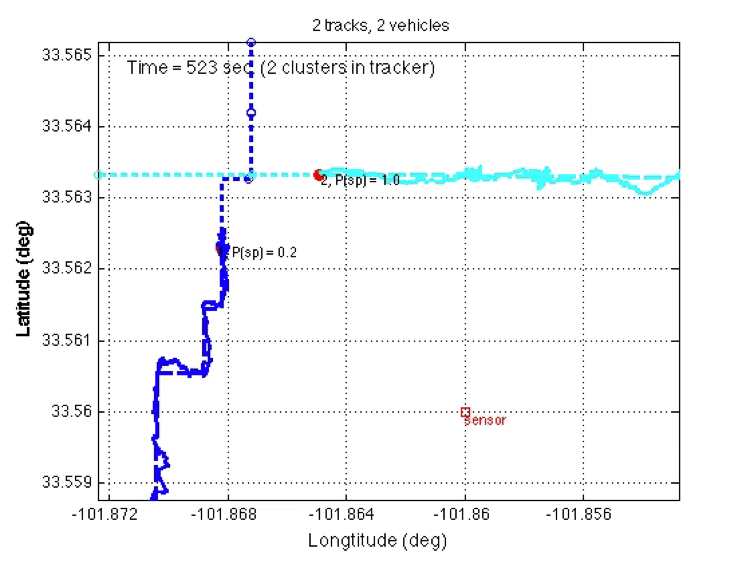}
\end{subfigure}

\caption{Trackers before intersection.}
\label{fig23}
\end{figure}

\begin{figure}[h]
\begin{subfigure}{0.475\textwidth}
\centering
\includegraphics[scale=0.5]{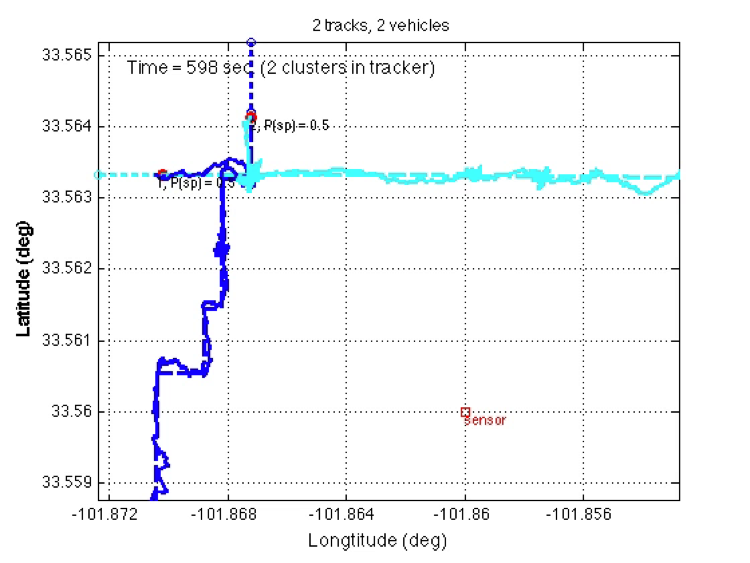}
\end{subfigure} %
\begin{subfigure}{0.475\textwidth}
\centering
\includegraphics[scale=0.5]{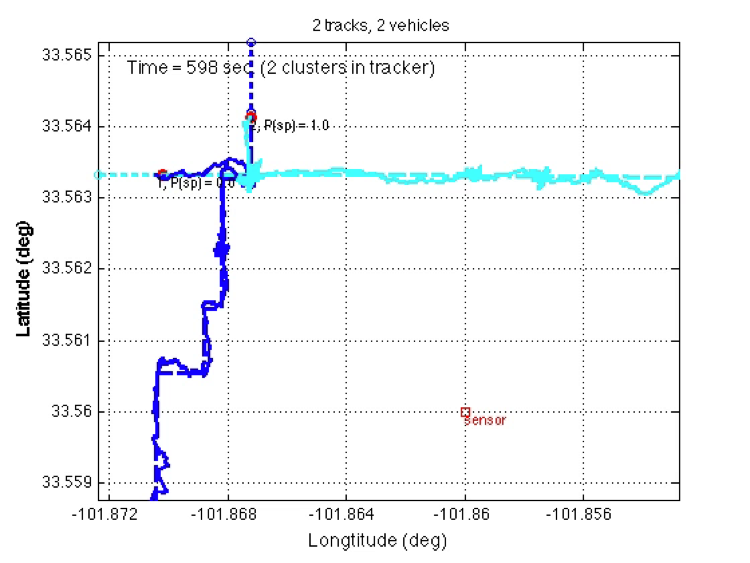}
\end{subfigure}

\caption{Trackers after intersection but before LLR update.}
\label{fig24}
\end{figure}

\begin{figure}[h]
\begin{subfigure}{0.475\textwidth}
\centering
\includegraphics[scale=0.5]{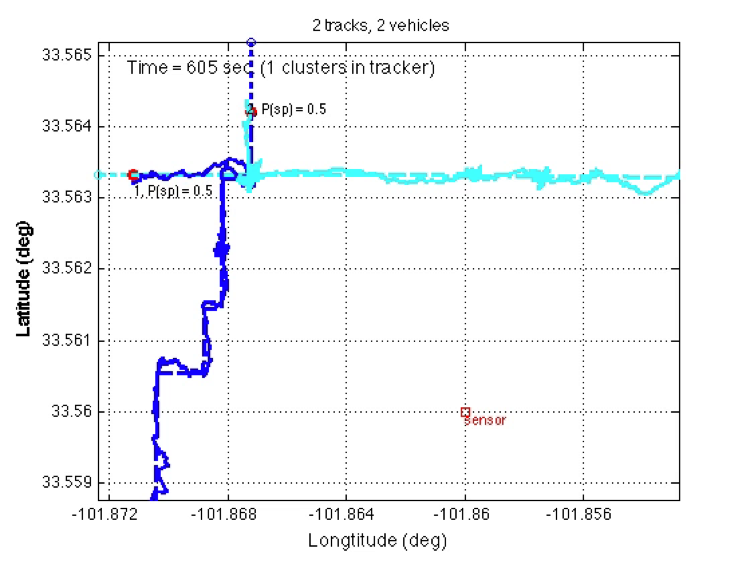}
\end{subfigure} %
\begin{subfigure}{0.475\textwidth}
\centering
\includegraphics[scale=0.5]{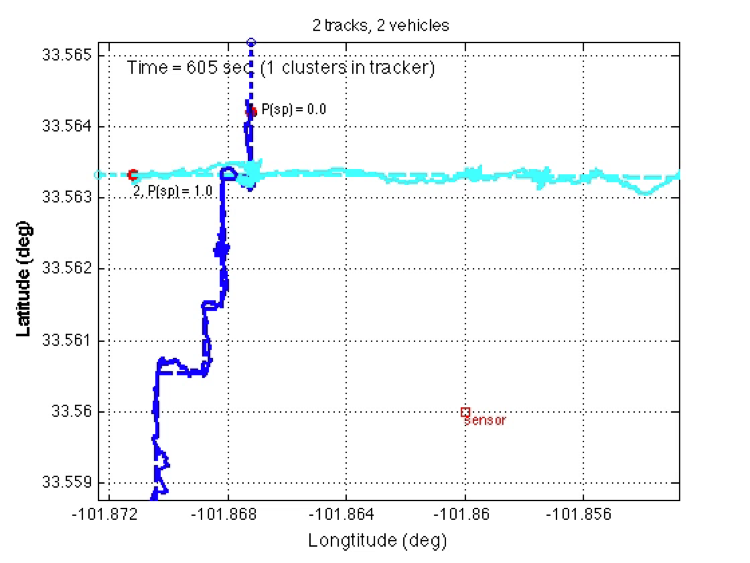}
\end{subfigure}

\caption{Trackers after intersection and LLR update.}
\label{fig25}
\end{figure}

Tracking is shown in Figures \ref{fig23} through \ref{fig25} using MHT with and without topological features. 
The MHT is run with a maximum of 10 multi-track hypotheses to provide the ability to carry enough 
alternative hypotheses to correct past association mistakes as new data clarifies association. 
The logistic regression approach described previously was used to perform machine learning to 
produce a behavior classifier. 
The target behavior classification provided by the topological features is shown as the probability of 
speeder  where a low probability indicates a normal driver and a high probability indicates a speeder. 
To emphasize that the tracker without topological features cannot classify driver behavior the 
probability of speeder without topological features is set to 0.5.

In Figure \ref{fig23}, the tracker results are shown prior to the intersection. 
The tracks are the same with and without the topological features because there is little confusion 
for these tracks before the intersection. 
But the tracker with topological features has accurate estimates of driving behavior. 
In Figure \ref{fig24}, the tracking picture is shown a short time after the intersection before enough time 
has passed for a topological feature update to the LLR score. 
Both trackers have the same incorrect tracks that switch vehicles at the intersection. 
It is in Figure \ref{fig25} that the difference occurs. 
MHT with topological features is able to correct the association error when it updates the 
LLR after the intersection. 
It estimated driver behavior before the intersection and as the behavior emerged after the intersection it could use the information from the behavior to correctly associate the data. 
Without topological features the tracker could not correct itself.

\begin{figure}[h]
\begin{center}
\includegraphics[scale=0.475]{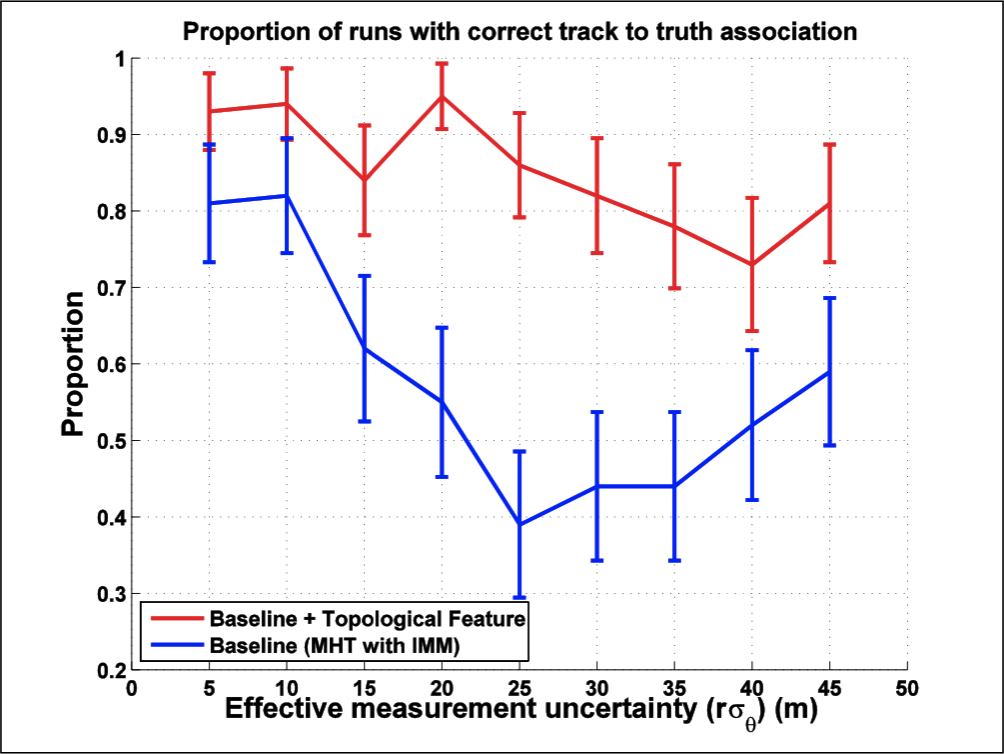}
\caption{Monte Carlo association performance as measurement uncertainty varies.}
\label{fig26}
\end{center}
\end{figure}

In order to establish the statistical performance with and without topological features, 
Monte Carlo results are shown in Figure \ref{fig26} for varying levels of measurement uncertainty.  
For each measurement uncertainty, 100 trials were performed and statistical error bars are shown. 
It is clear that association is better with topological features. 
In fact, topological features allow tracking to be successful at substantially higher levels of measurement noise. 

\section{Conclusion}

\label{sec:conclusion}

We have shown that persistence, combined with learning, provides an effective tool to measure 
agent activity and we have demonstrated how to apply that measure to improve image based tracking.
There are a large number of ways that this methodology can be generalized, including the use of more 
sophisticated measurement functions of driving behavior. 
For example, we tested the use of delay 
reconstruction to transform functions into point clouds - the results can then be analyzed with a variety
of data analysis tools.
Whatever measure is used, one can train the weighting of topological features on different environments,
including different cities, rural or mountainous environments, etc.
And one can include other types of intelligence such as cell-phone signatures, tweets, etc. to improve
tracking performance. 
It is clear that topology + learning + multi-Int provides a wealth of new approaches to behavioral description
and tracking.

Another important generalization of the ideas in this paper is the capture of {\em collaborative
behavior} among a collection of agents.
This behavior is more likely to exhibit itself at the large scale, as drivers act in parallel and make 
interesting driving patterns (departing from the same point and remerging later, etc.).
In many ways, this is what persistent topology should be best at, given that it looks at large scale 
behavior rather than the fine details, yet still provides quantitative measurement of behavioral patterns.
To do this will require  the use of higher dimensional persistence, so this is beyond the scope
of the current paper.
We have already explored the use of
one-dimensional persistence to track ``inefficient driving'' or ``loopiness''
\cite{Munch2013}, stopping times, and other behaviors that indicate nonchalance.
Many other such patterns are detectable in this way.


%

\bibliography{Paul.bib}{}
\bibliographystyle{plain}

\section*{Acknowledgment}

This work was partially supported by the grants OASIS JHU Subcontract 110279 on HQ0034-12-C-0024, 
DARPA LIMS Subcontract on HDTRA1-11-1-0048, and NSF-DMS grant 10-45153.

\appendix[Additional Figures]

\begin{figure}[h]
\begin{center}
\includegraphics[width=0.4\textwidth]{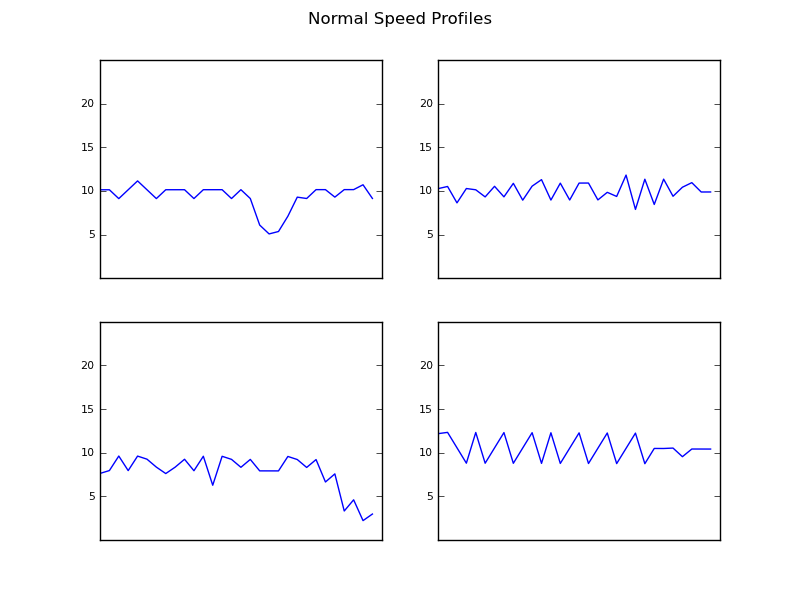}
\caption{Speed profiles of normal drivers.}
\label{fig:norm_spds}
\end{center}
\end{figure}

\begin{figure}[h]
\begin{center}
\includegraphics[width=0.4\textwidth]{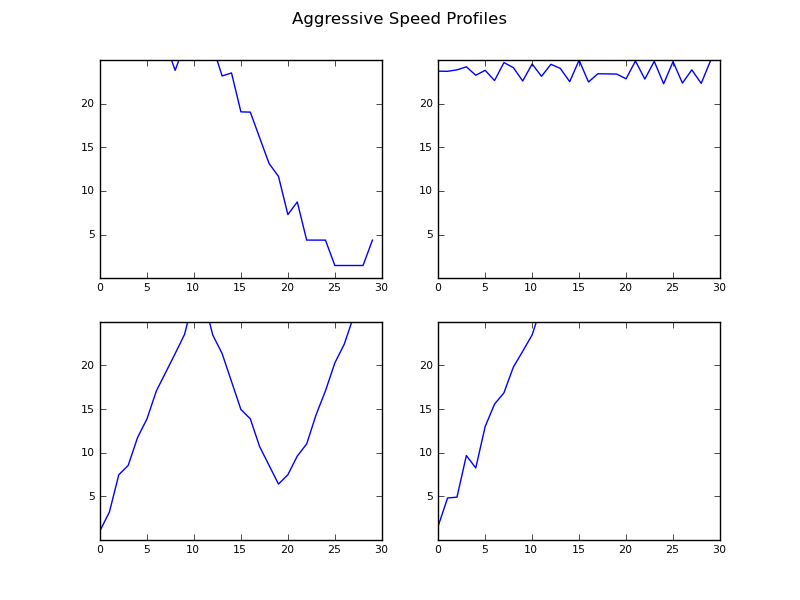}
\caption{Speed profiles of aggressive drivers.}
\label{fig:agg_spds}
\end{center}
\end{figure}

\begin{figure}[h]
\begin{center}
\includegraphics[width=0.4\textwidth]{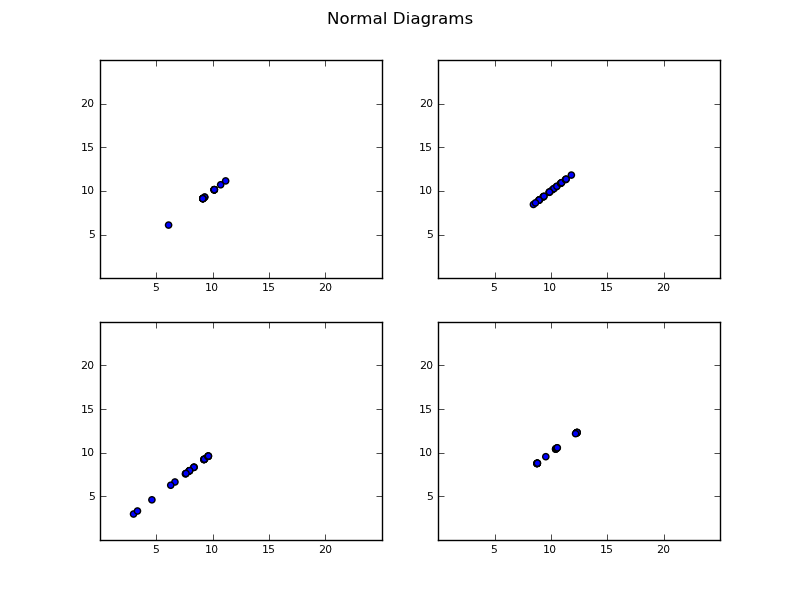}
\caption{Augmented diagrams for the normal speed profiles.}
\label{fig:norm_diagrams}
\end{center}
\end{figure}

\begin{figure}[h]
\begin{center}
\includegraphics[width=0.4\textwidth]{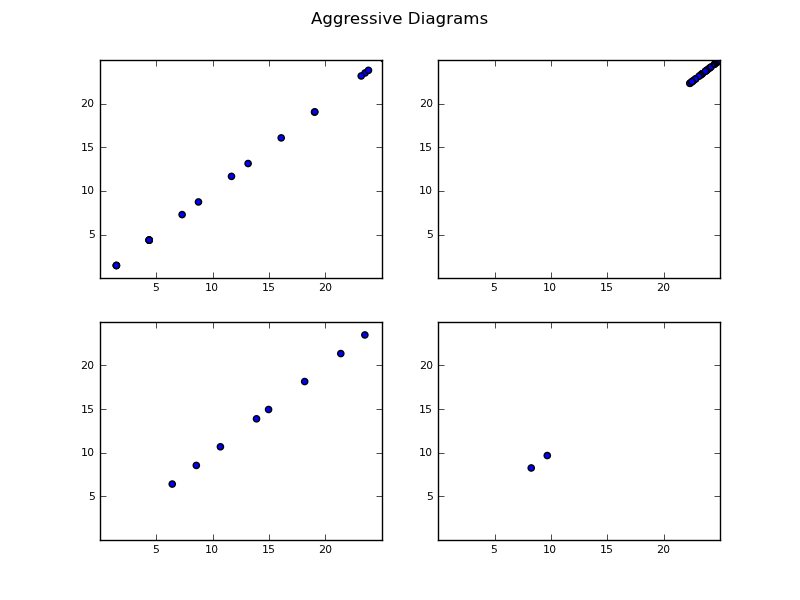}
\caption{Augmented diagrams for the aggressive speed profiles.}
\label{fig:agg_diagrams}
\end{center}
\end{figure}

\begin{figure}[h]
\begin{center}
\includegraphics[width=0.4\textwidth]{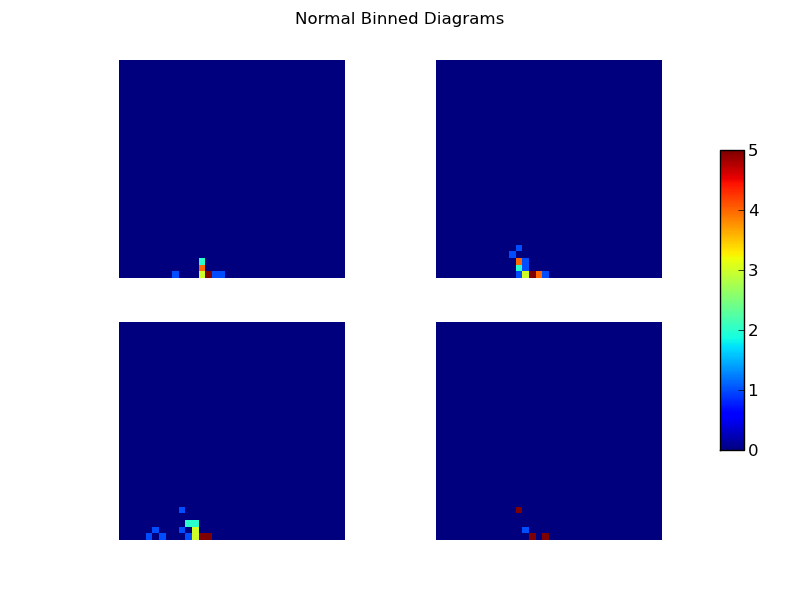}
\caption{Binned diagrams for the normal speed profiles.}
\label{fig:norm_bin_diagrams}
\end{center}
\end{figure}

\begin{figure}[h]
\begin{center}
\includegraphics[width=0.4\textwidth]{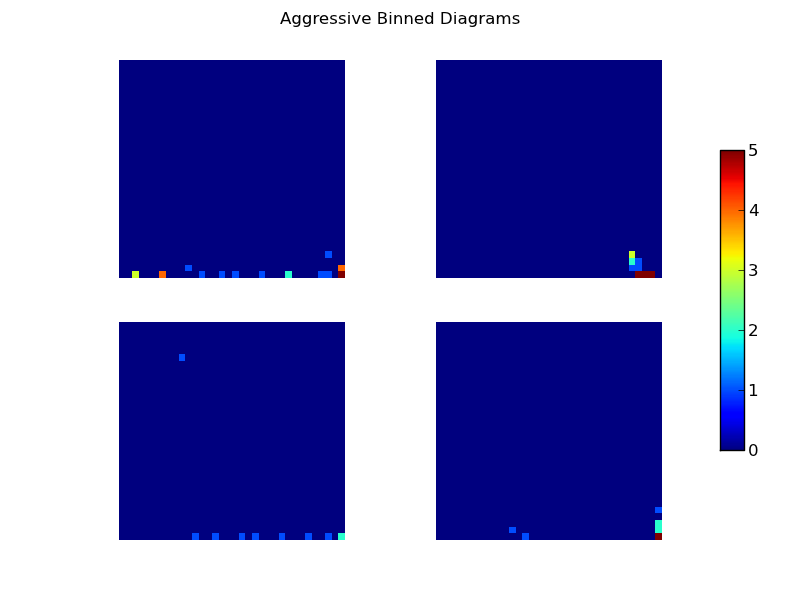}
\caption{Binned diagrams for the aggressive speed profiles.}
\label{fig:agg_bin_diagrams}
\end{center}
\end{figure}

\begin{figure}[h]
\begin{center}
\includegraphics[width=0.4\textwidth]{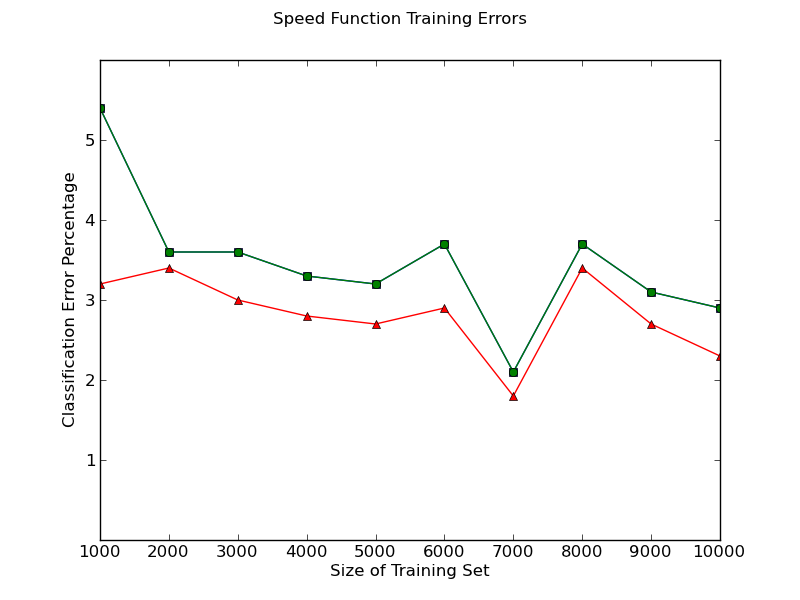}
\caption{Test errors using binned diagrams computed from the speed function. The test errors of the logistic regression (red triangles), the Poisson model (blue squares), and the multinomial model (green circles) as a function of the size of the training data.}
\label{fig:spd_err}
\end{center}
\end{figure}

\begin{figure}[h]
\begin{center}
\includegraphics[width=0.4\textwidth]{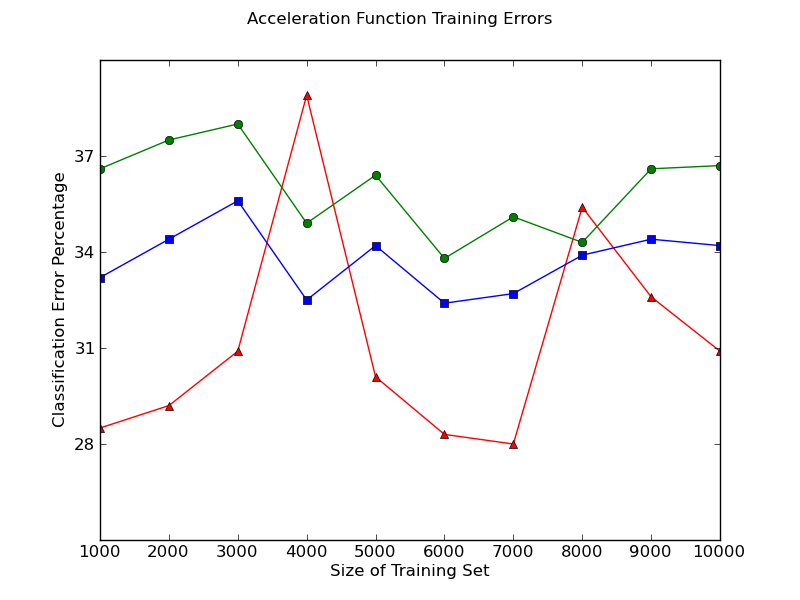}
\caption{Test errors using binned diagrams computed from the acceleration function. The test errors of logistic regression (red triangles), the Poisson model (blue squares), and the multinomial model (green circles) as a function of the size of the training data.}
\label{fig:acc_err}
\end{center}
\end{figure}

\begin{figure}[h]
\begin{center}
\includegraphics[width=0.4\textwidth]{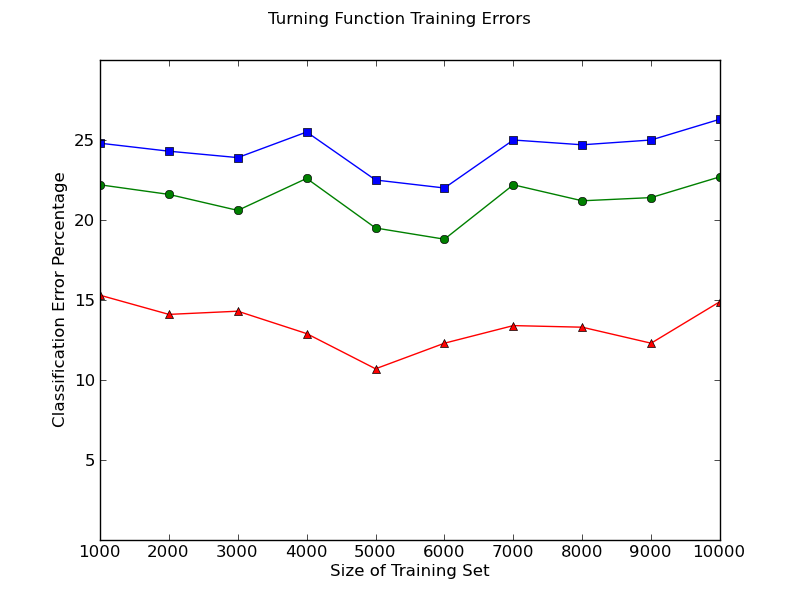}
\caption{Test errors using binned diagrams computed from the turning function. The test errors of logistic regression (red triangles), the Poisson model (blue squares), and the multinomial model (green circles) as a function of the size of the training data.}
\label{fig:trn_err}
\end{center}
\end{figure}

\begin{figure}[h]
\begin{center}
\includegraphics[width=0.4\textwidth]{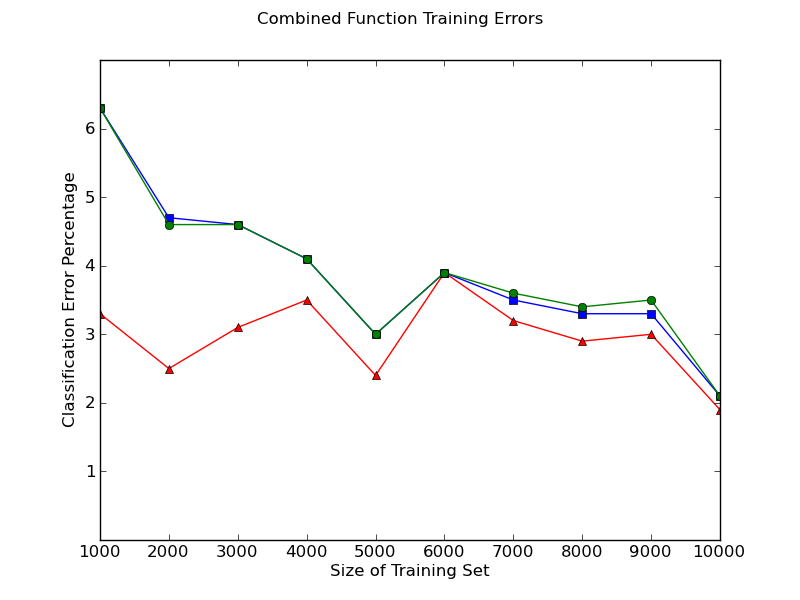}
\caption{Test errors using binned diagrams computed from the the speed, acceleration, and turning functions all combined. The test errors of logistic regression (red triangles), the Poisson model (blue squares), and the multinomial model (green circles) as a function of the size of the training data.}
\label{fig:cmb_err}
\end{center}
\end{figure}

\begin{figure}[h]
\begin{center}
\includegraphics[width=0.4\textwidth]{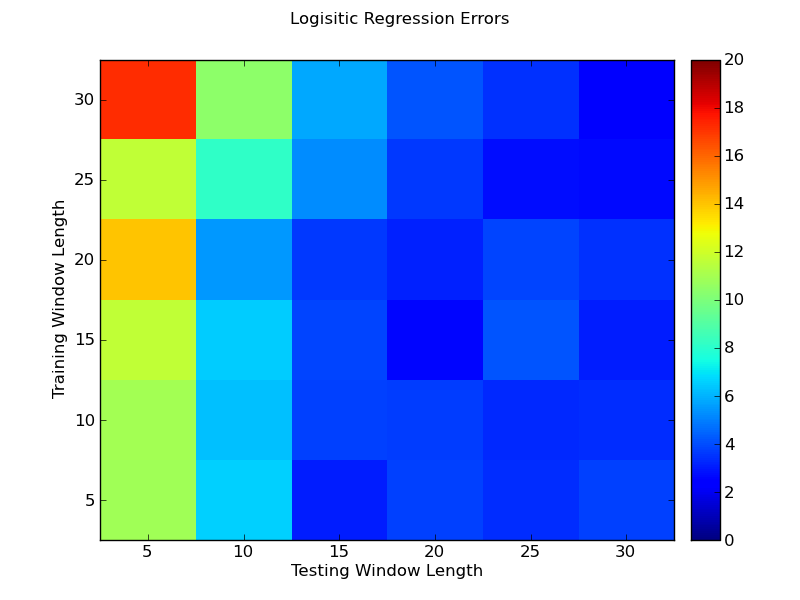}
\caption{Average classification percentage errors for logistic regression over 25 randomized training runs.}
\label{fig:LRerr}
\end{center}
\end{figure}

\begin{figure}[h]
\begin{center}
\includegraphics[width=0.4\textwidth]{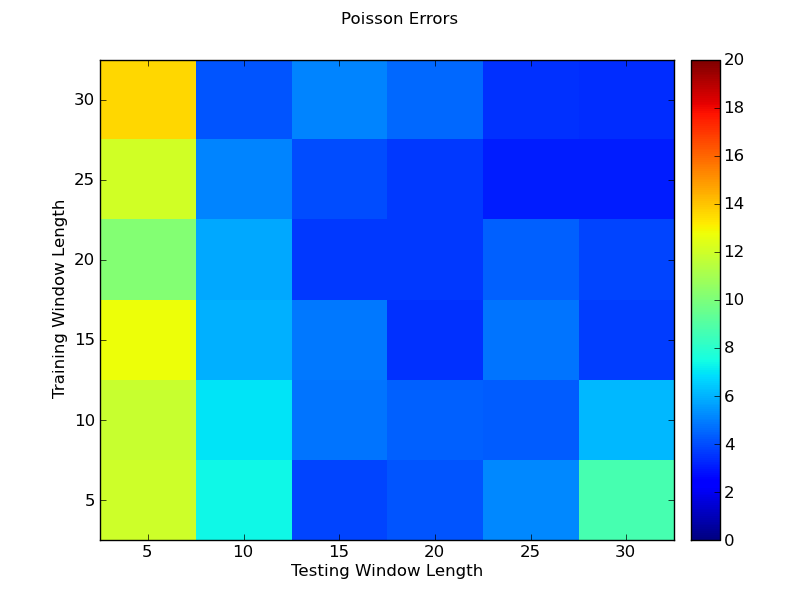}
\caption{Average classification percentage errors for classification under a Poisson model, with randomized input order over 25 runs.}
\label{fig:Perr}
\end{center}
\end{figure}

\begin{figure}[h]
\begin{center}
\includegraphics[width=0.4\textwidth]{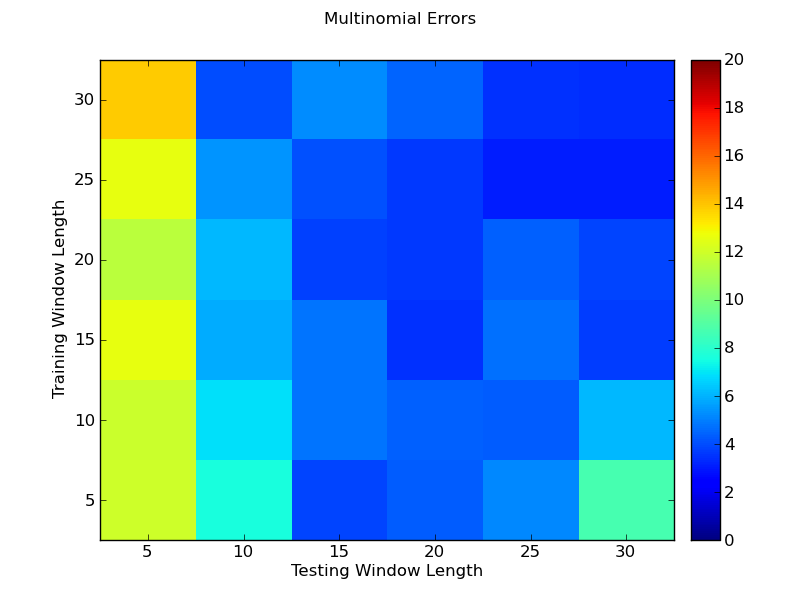}
\caption{Average classification percentage errors for classification under a multinomial model, with randomized input order over 25 runs.}
\label{fig:Merr}
\end{center}
\end{figure}

\begin{figure}[h]
\begin{center}
\includegraphics[width=0.4\textwidth]{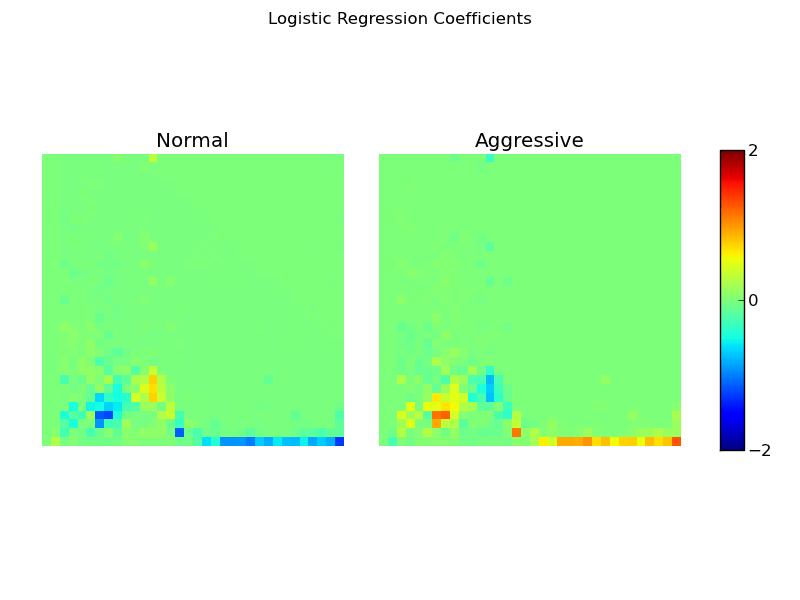}
\caption{Learned logistic regression coefficients.}
\label{fig:lrcoeff}
\end{center}
\end{figure}

\begin{figure}[h]
\begin{center}
\includegraphics[width=0.4\textwidth]{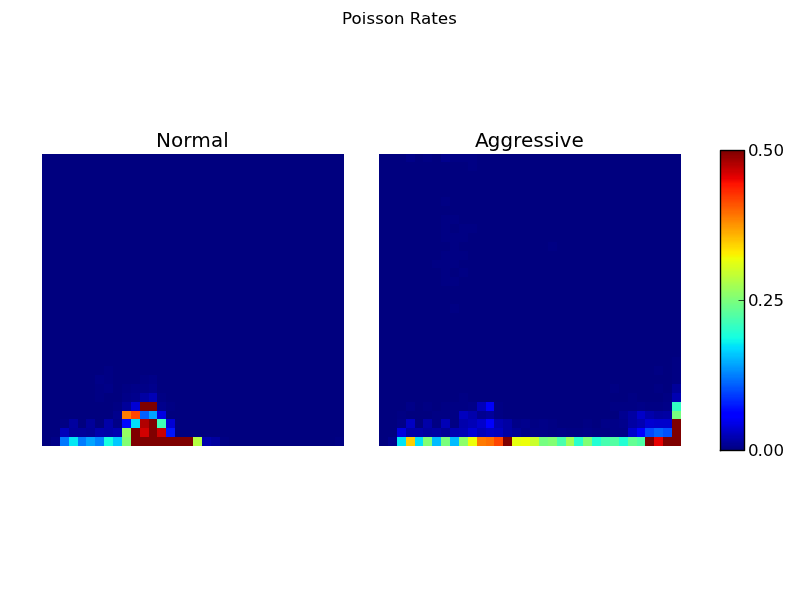}
\caption{Learned Poisson rates.}
\label{fig:prates}
\end{center}
\end{figure}

\begin{figure}[h]
\begin{center}
\includegraphics[width=0.4\textwidth]{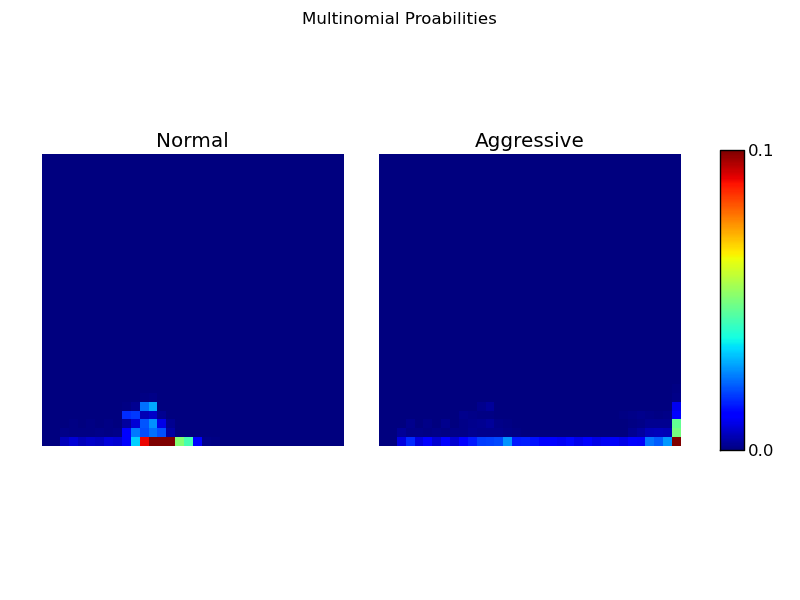}
\caption{Learned multinomial probabilities.}
\label{fig:mprobs}
\end{center}
\end{figure}


\ifCLASSOPTIONcaptionsoff
  \newpage
\fi

\end{document}